\documentclass[prd,amsmath,amssymb,superscriptaddress,preprintnumbers,nofootinbib,twocolumn,10pt]{revtex4}
\pdfoutput=1
\usepackage{graphicx}
\usepackage{dcolumn}
\usepackage{bm}
\usepackage{amssymb}
\usepackage{latexsym}
\usepackage{booktabs}
\usepackage{amsmath}
\usepackage{multirow}
\usepackage{url}
\usepackage{float}
\usepackage[colorlinks=true, linkcolor=red, citecolor=blue]{hyperref}

\newcommand{\ve}{\varepsilon}

\usepackage[normalem]{ulem}
\usepackage{color}
\usepackage{array}
\usepackage{enumerate}

\begin{document}

\title{A comprehensive forecast for cosmological parameter estimation using joint observations of gravitational waves and short $\gamma$-ray bursts}

\author{Tao Han}
\affiliation{Key Laboratory of Cosmology and Astrophysics (Liaoning) \& College of Sciences, Northeastern University, Shenyang 110819, China}
\author{Shang-Jie Jin}
\affiliation{Key Laboratory of Cosmology and Astrophysics (Liaoning) \& College of Sciences, Northeastern University, Shenyang 110819, China}
\author{Jing-Fei Zhang}
\thanks{jfzhang@mail.neu.edu.cn (corresponding author)}
\affiliation{Key Laboratory of Cosmology and Astrophysics (Liaoning) \& College of Sciences, Northeastern University, Shenyang 110819, China}
\author{Xin Zhang}
\thanks{zhangxin@mail.neu.edu.cn (corresponding author)}
\affiliation{Key Laboratory of Cosmology and Astrophysics (Liaoning) \& College of Sciences, Northeastern University, Shenyang 110819, China}
\affiliation{Key Laboratory of Data Analytics and Optimization for Smart Industry (Ministry of Education), Northeastern University, Shenyang 110819, China}
\affiliation{National Frontiers Science Center for Industrial Intelligence and Systems Optimization, Northeastern University, Shenyang 110819, China}

\begin{abstract}

In the third-generation (3G) gravitational-wave (GW) detector era, the multi-messenger GW observation for binary neutron star (BNS) merger events can exert great impacts on exploring the cosmic expansion history. In this work, we comprehensively explore the potential of 3G GW standard siren observations in cosmological parameter estimations by considering the 3G GW detectors and the future short $\gamma$-ray burst (GRB) detector THESEUS-like telescope joint observations. Based on the 10-year observation of different detection strategies, we predict that the numbers of detectable GW-GRB events are 334--674 with the redshifts $z<3.5$ and the inclination angles $\iota<15^{\circ}$. For the cosmological analysis, we consider the $\Lambda$CDM, $w$CDM, $w_0w_a$CDM models, and interacting dark energy (IDE) models. We find that GW can tightly constrain the Hubble constant with precisions of 0.345\%--0.065\%, but perform not well in constraining other cosmological parameters. Fortunately, GW could effectively break the cosmological parameter degeneracies generated by the mainstream EM observations, CMB+BAO+SN (CBS). When combining the mock GW data with the CBS data, CBS+GW can tightly constrain the equation of state of dark energy $w$ with a precision of 1.26\%, close to the standard of precision cosmology. Meanwhile, the addition of GW to CBS could improve constraints on cosmological parameters by 34.2\%--94.9\%. In conclusion, GW standard siren observations from 3G GW detectors could play a crucial role in helping solve the Hubble tension and probe the fundamental nature of dark energy.

\end{abstract}

\maketitle

\section{Introduction}

In 1998, the accelerated expansion of the universe was discovered by type Ia supernovae (SN) observations~\cite{Riess:1998cb,Perlmutter:1998np}. 
The cosmic acceleration is usually explained by assuming an exotic component with negative pressure, known as dark energy~\cite{Peebles:2002gy,Copeland:2006wr,Li:2011sd}.
In recent years, the $\Lambda$ cold dark matter ($\Lambda$CDM) model, which is widely viewed as the standard model of cosmology, has been established.
The measurements of cosmic microwave background (CMB) anisotropies by the $Planck$ mission strongly favor a six-parameter base $\Lambda$CDM model, in which dark energy is served by the cosmological constant $\Lambda$. However, recent advancements in the precision of cosmological parameter measurements have revealed some cracks in the $\Lambda$CDM model. Notably, it is found that the tension between the values of the Hubble constant inferred from the CMB observation (assuming the $\Lambda$CDM model)~\cite{Planck:2018vyg} and the Cepheid-supernova distance ladder measurement (model-independent)~\cite{Riess:2021jrx} has now been in more than 5$\sigma$ \cite{Riess:2021jrx}. Recently, the Hubble tension has been intensively discussed in the literature~\cite{Li:2013dha,Zhao:2017urm,cai:2020,Guo:2018ans,Guo:2019dui,Yang:2018euj,Vagnozzi:2019ezj,DiValentino:2019jae,DiValentino:2019ffd,Feng:2019jqa,Lin:2020jcb,Li:2020tds,Hryczuk:2020jhi,Gao:2021xnk,Cai:2021wgv,Vagnozzi:2021tjv,Vagnozzi:2021gjh}. The Hubble tension is now commonly considered a severe crisis for cosmology \cite{Riess:2019qba,Verde:2019ivm}. On one hand, the tension may herald the possibility of new physics beyond the standard $\Lambda$CDM model. Nevertheless, no consensus has been reached on a valid extended cosmological model that can truly solve the tension. On the other hand, the methods that can independently measure the Hubble constant need to be greatly developed to make an arbitration for the Hubble tension. The gravitational wave (GW) standard siren method is one of the most promising methods.

As proposed by Schutz in 1986, GW can be used as standard siren to explore the cosmic expansion history~\cite{Schutz:1986gp,Holz:2005df}. The absolute luminosity distance to the source could be directly obtained through the analysis of the GW waveform. If the redshift of the source can also be obtained by identifying its electromagnetic (EM) counterpart (usually referred to as bright siren) or using statistical methods to infer the redshift information (usually referred to as dark siren), then we can establish the true distance-redshift relation for exploring the expansion history of the universe; see e.g., Refs.~\cite{Zhao:2010sz,Cai:2017aea,Cai:2016sby,Wang:2018lun,Zhang:2019ylr,Zhang:2019loq,Zhang:2018byx,Li:2019ajo,Jin:2020hmc,Zhang:2019ple,Jin:2023zhi,Jin:2021pcv,Wu:2022dgy,Jin:2022tdf,Jin:2023tou,Wang:2019tto,Wang:2021srv,Zhao:2019gyk,Jin:2023sfc,Hou:2022rvk,Li:2023gtu,Qi:2021iic,Wang:2022oou,Dong:2024bvw,Feng:2024lzh,Song:2022siz,KAGRA:2013rdx,2022SCPMA..6510431G}. So far, the only bright siren event GW170817 has given the first measurement of the Hubble constant using the standard siren method with a precision of about 14\% \cite{LIGOScientific:2017adf}. In addition, the dark siren method gives a 19\% measurement precision of the Hubble constant \cite{LIGOScientific:2021aug}. The fact means that the current measurements are far from making an arbitration for the Hubble tension and one would have to resort to the future GW observations. 

The third generation (3G) ground-based GW detectors, the Einstein Telescope (ET)~\cite{ET-web,Punturo:2010zz} and the Cosmic Explorer (CE)~\cite{CE-web,Evans:2016mbw} show a powerful sensitivity, which is more than one order of magnitude improved over the current detectors \cite{Evans:2021gyd}. Thus, in the era of 3G GW detectors, more binary neutron star (BNS) mergers will be detected at much deeper redshifts~\cite{Jin:2022qnj}. However, the detection of EM counterparts is still difficult. The observed BNS merger rate, as inferred by the LIGO-Virgo-KAGRA collaboration, ranges from $10$ to $1700~\mathrm{Gpc^{-3}~yr^{-1}}$ \cite{KAGRA:2021duu}, implying that close events similar to GW170817 may occur approximately once a decade.
What's more, a single GW detector, even in a triangular configuration as planned for ET, has limited localization capabilities, giving a much larger number of galaxies per error region~\cite{Chen:2020zoq}. Fortunately, a fraction of BNS mergers (those viewed roughly on-axis) are expected to be accompanied by short $\gamma$-ray bursts (GRBs) and their associated afterglows, which can be accurately localized by the GRB detectors. Consequently, GRB detection plays a vital role in facilitating the subsequent identification of host galaxies and the determination of redshifts.

Due to the strong beaming nature of short GRBs \cite{Howell:2018nhu}, only GRBs with small inclination angles are detectable. Recent works show that only about 0.1\% BNS mergers have detectable EM counterparts \cite{Yu:2021nvx}. However, comprehensive forecasts for cosmological parameter estimations with 3G era GW-GRB joint observations are still absent, which deserves an in-depth investigation. In this work, we focus on the synergy of 3G GW detectors with a future GRB detector with the characteristic of the proposed Transient High-Energy Sky and Early Universe Surveyor (THESEUS) mission~\cite{THESEUS:2017qvx,THESEUS:2017wvz,Stratta:2018ldl}. We constrain five typical cosmological models including the $\Lambda$CDM, $w$CDM, $w_0w_a$CDM, and interacting dark energy (IDE) models (I$\Lambda$CDM and I$w$CDM). Here we highlight three points upgraded in this paper: (i) We conduct a comprehensive and robust analysis of GW-GRB detection and calculate the redshift distribution of GW-GRB events instead of assuming 1000 detected standard sirens in 10-year observation, as adopted in Refs.~\cite{Zhao:2010sz,Cai:2017aea,Cai:2016sby,Wang:2018lun,Zhang:2019ylr,Zhang:2019loq,Zhang:2018byx,Li:2019ajo,Jin:2020hmc,Zhang:2019ple,Jin:2023zhi,Jin:2021pcv,Wu:2022dgy,Jin:2022tdf}. (ii) For the 3G GW detectors, the impact of the Earth's rotation is important, which includes two effects: one is the modulation of the Doppler effect quantified by the time-dependent function $\Phi_{ij}$, and the other is quantified by the time-dependent detector responses $F_{+,k}$ and $F_{\times,k}$. Therefore, we take into account the Earth's rotation in the simulation of GW standard sirens, bringing it closer to real observations. (iii) We make an analysis of the optimistic and realistic scenarios for GW-GRB detection and perform cosmological analysis, which could better show the potential of the cosmological parameter estimation using the 3G era standard sirens.

Recently, Hou~\emph{et al.}~\cite{Hou:2022rvk} constrained IDE models with future GW and GRB joint observation. However, they only focused on the synergy of ET alone with the THESEUS mission in the optimistic scenario for GW-GRB detection. In this paper, we make a comprehensive analysis of four different cases of 3G GW observations, single ET, single CE, the CE-CE network, and the ET-CE-CE network by using the Fisher information matrix. Moreover, we also consider the realistic scenario for GW-GRB detection. For IDE models, we employ the extended parameterized post-Friedmann (ePPF) approach~\cite{Li:2014eha,Li:2014cee,Li:2015vla,Li:2023fdk} to avoid the cosmological perturbations (see Sec.~\ref{sec5} for a detailed discussion).

This paper is organized as follows. In Sec.~\ref{sec2} we introduce the method to simulate the GW standard siren data. In Sec.~\ref{sec3}, we present the results of the prediction for GW-GRB detection. In Sec.~\ref{sec4}, we discuss the impact of the Earth’s rotation to the luminosity distance uncertainties. In Sec.~\ref{sec5}, we show the constraint results of cosmological parameters from the GW-GRB observations. The conclusion is given in Sec.~\ref{sec6}. Throughout this paper, the fiducial values of cosmological parameters are set to the constraint results from CMB ({\it Planck} 2018 TT, TE, EE+lowE), baryon acoustic oscillation (BAO), and SN data. Unless otherwise specified, we set $G = c = 1$.

\section{Methodology}\label{sec2}

\subsection{Cosmological models}

In a flat Friedmann-Roberston-Walker universe, we can obtain the energy conservation equations,
\begin{align}
	&\dot{\rho}_{\rm de} +3H(1+w)\rho_{\rm de}= Q,\\
	&\dot{\rho}_{\rm c} +3H\rho_{\rm c}=-Q,
\end{align}
where $Q$ is the energy transfer rate, $\rho_{\rm de}$ and $\rho_{\rm c}$ represent the energy densities of dark energy and CDM, respectively, $w$ is the equation of state (EoS) parameter of dark energy, $H$ is the Hubble parameter, and the dot denotes the derivative with respect to the cosmic time $t$. 

If $Q=0$, it indicates no interaction between dark energy and CDM. In this work, we consider three cosmological models without interaction: (i) $\Lambda$CDM model, known as the standard model of cosmology, in which dark energy is described by a cosmological constant $\Lambda$ with $w(z)=-1$; (ii) $w$CDM model, the simplest dynamical dark energy model, in which the EoS parameter of dark energy is a constant, i.e., $w(z)=\rm {constant}$; (iii) $w_0w_a$CDM model, the parameterized dynamical dark energy model with $w(z)=w_0+w_{\rm a}z/(1+z)$.

If $Q\neq0$, it means that dark energy has a direct interaction with CDM. This type of cosmological model is referred to as an IDE model.
However, the microscopic nature of dark energy and dark matter is still unclear, the energy transfer rate in the IDE models can be considered in a purely phenomenological way, i.e., proportional to the energy density of dark energy, dark matter, or some mixture of them. In this work, we will consider the interaction form of $Q=\beta H\rho_{\rm c}$, where $\beta$ is the dimensionless coupling parameter. Here $\beta>0$ and $\beta<0$ mean CDM decaying into dark energy and dark energy decaying into CDM, respectively.

\subsection{Simulation of BNS mergers}

In order to generate a catalog of BNS coalescences, we first calculate a redshift distribution of the BNS mergers. It is drawn from a normalized probability distribution,
\begin{equation}
	p(z)=\frac{R_{\rm m}(z)}{\int_{0}^{10}R_{\rm m}(z)dz},
\end{equation}
where $R_{\rm m}(z)$ is the BNS merger rate with redshift $z$ in the observer frame. It can be estimated by
\begin{equation}
	R_{\rm m}(z)=\frac{\mathcal{R}_{\rm m}(z)}{1+z} \frac{dV(z)}{dz},
\end{equation}
where $dV/dz$ is the comoving volume element and $\mathcal{R}_{\rm m}(z)$ is the BNS merger rate in the source frame.

BNS merger can be thought as occurring with a delay timescale with respect to the BNS formation history, which is given by
\begin{equation}
	\mathcal{R}_{\rm m}(z)=\int_{t_{\rm min}}^{t_{\rm max}} \mathcal{R}_{\rm f}[t(z)-t_{\rm d}] P(t_{\rm d})dt_{\rm d},
\end{equation}
where $P(t_{\rm d})$ is the delay time distribution which encodes the time span between the formation of the BNS system until the two NSs merge through the emission of GWs and GRBs, $t_{\rm d}$ is the time delay between the formation of BNS system and merger, $t(z)$ is the age of the universe at the time of merger, $t_{\rm min}=20$ Myr is the minimum delay time, $t_{\rm max}=t_{\rm H}$ is the Hubble time which stands for a maximum delay time~\cite{Belgacem:2019tbw}, and $\mathcal{R}_{\rm f}$ is the BNS formation rate.

Here, we assume that $\mathcal{R}_{\rm f}$ is simply proportional to the star formation rate density:
\begin{equation}
	\mathcal{R}_{\rm f}\equiv \lambda \psi_{\rm {MD}},
\end{equation}
where $\psi_{\rm {MD}}$ is the Madau-Dickinson star formation rate~\cite{Madau:2014bja} and $\lambda$ is the currently unknown BNS mass efficiency (assumed not to evolve with redshift) used as a free parameter~\cite{Safarzadeh:2019pis}, determined by the local comoving merger rate $\mathcal{R}_{\rm m}(z=0)$.

Main types of delay time distributions include the Gaussian delay model~\cite{Virgili:2009ca}, log-normal delay model~\cite{Nakar:2005bs,Wanderman:2014eza}, exponential time delay model~\cite{Sathyaprakash:2019rom,Vitale:2018yhm}, and power-law delay model~\cite{Virgili:2009ca,DAvanzo:2014urr}. For simplicity, we only adopt the exponential time delay model as our delay time model, which is given by \cite{Sathyaprakash:2019rom,Vitale:2018yhm}
\begin{equation}
	P(t_{\rm d})=\frac{1}{\tau}{\rm exp}(-t_{\rm d}/\tau),
\end{equation}
with $\tau=100$ Myr for $t_{\rm d}>t_{\rm min}$.

In our calculations, we adopt the local comoving merger rate to be 920 $\rm Gpc^{-3}~yr^{-1}$ estimated from the O1 LIGO and the O2 LIGO/Virgo observation run~\cite{LIGOScientific:2018mvr}. This is also consistent with the latest O3 observation run~\cite{KAGRA:2021duu}. We simulate a catalog of BNS mergers for 10 years. For each source, the location $(\theta,\phi)$, the cosine of the inclination angle $\iota$, the polarization angle $\psi$, and the coalescence phase $\psi_{\rm c}$ are drawn from uniform distributions. For the masses in a BNS system, we assume that each component is drawn independently from a common Gaussian distribution, based on the NS masses observed in Galactic BNSs. This distribution has a mean of 1.33 $M_{\odot}$ and a standard deviation of 0.09 $M_{\odot}$~\cite{Ozel:2016oaf,LIGOScientific:2018mvr}.


It should be noted that when simulating the catalog of BNS coalescences, we randomly select relevant parameters within reasonable ranges consistent with the O3 observing run~\cite{KAGRA:2021duu}. Therefore, we must admit that it is challenging to fully eliminate the influence of significant stochastic fluctuations on the constraints on cosmological parameters. The primary goal of this paper is to illustrate the significant potential that lies in the collaborative observations between 3G GW detectors and future GRB detectors in refining our understanding of cosmological researches.

\subsection{Detection of GW events}

In this section, we briefly review the detection of GW detector network. We use the vector $r_k$ with $k=1,2,\ldots,N$ to denote the spatial locations of the detectors, which is given by

\begin{equation}
\boldsymbol{r}_k=R_{\oplus}(\sin\varphi_k\cos\alpha_k,\sin\varphi_k\sin\alpha_k,\cos\varphi_k),
\end{equation}
where $R_{\oplus}$ is the radius of the Earth, $\varphi_k$ is the latitude of the detector in the celestial system. Here we define $\alpha_k$ as $\alpha_k\equiv\lambda_k+\Omega_{\rm r}t$, where $\lambda_k$ is the East longtitude of the detector, $\Omega_{\rm r}$ is the rotational angular velocity of the Earth. In this paper, we take the zero Greenwich sidereal time at $t=0$.

Let us begin by considering the long-wavelength approximation. Under this approximation, the antenna response is purely a projection effect that maps the strains in the wave-frame onto the detector~\cite{Jaranowski:1998qm}.\footnote{Here we ignore the errors due to the long-wavelength approximation optimistically. We assume the associated GW wavelengths are much longer than the detector’s arms and neglect the frequency dependence of detectors’ responses. In fact, the interferometric GW detectors are dynamic instruments. The antenna response functions generally depend on the frequency of GW (see e.g., Ref.~\cite{Essick:2017wyl} for more discussions). The details of the errors due to the long-wavelength approximation will be investigated in our future work.} For a transverse-traceless GW signal detected by a single detector labeled by $k$, the response is given by
\begin{equation}
h_k(t_{0}+\boldsymbol{\tau}_k+t)=F_{+,k}h_{+}(t)+F_{\times,k}h_{\times}(t), 0<t<T,
\end{equation}
where $t_{0}$ is the arrival time of the GW at the coordinate origin, $\boldsymbol{\tau}_k=\boldsymbol{n}\cdot \boldsymbol{r}_k(t)$ is the time required for the GW to travel from the origin to reach the $k$-th detector at time $t$. Here $t\in [0,T]$ is the time label of GW, $T$ is the signal duration, and $\boldsymbol{n}$ is the propagation direction of a GW event. As mentioned above, $h_{+}$ and $h_{\times}$ are the plus and cross modes of GW, respectively. The quantites $F_{+,k}$ and $F_{\times,k}$ are the $k$-th detector's antenna response functions of two polarizations, which depend on the location of the source, the polarization angle, and the specific geometry parameters of the GW detector (the latitude $\varphi$, the longitude $\lambda$ of the detector’s vertex, the opening angle $\zeta$ between the detector’s two arms, and the orientation angle $\gamma$ of the detector’s arms measured counter-clockwise from East to the bisector of the interferometer arms)~\cite{Jaranowski:1998qm}.

Under the stationary phase approximation (SPA), the frequency-domain GW waveform considering the detector network including $N$ independent detectors can be written as~\cite{Wen:2010cr}
\begin{equation}
\tilde{\boldsymbol{h}}(f)=e^{-i\boldsymbol\Phi}\boldsymbol h(f),
\label{eq:waveform}
\end{equation}
where $\boldsymbol\Phi$ is the $N\times N$ diagonal matrix with $\Phi_{ij}=2\pi f\delta_{ij}(\boldsymbol{n\cdot r}_i(f))$, and

\begin{widetext}
\begin{equation}
\begin{aligned}
\boldsymbol h(f)=&\left [\frac{h_1(f)}{\sqrt{S_{\rm {n},1}(f)}}, \frac{h_2(f)}{\sqrt{S_{\rm {n},2}(f)}}, \ldots,\frac{h_k(f)}{\sqrt{S_{{\rm n},k}(f)}},\ldots,\frac{h_N(f)}{\sqrt{S_{{\rm n},N}(f)}}\right ]^{\rm T},
\end{aligned}
\end{equation}
\end{widetext}
where $h_k(f)$ is the frequency domain GW waveform and $S_{{\rm n},k}(f)$ is the one-side noise power spectral density of the $k$-th detector. In this paper, we consider the waveform in the inspiralling stage for a BNS system and neglect the NS spins, which are believed to have small effect on the GW waveform of BNS systems~\cite{OShaughnessy:2006uzj}. We adopt the restricted Post-Newtonian approximation and calculate the waveform to the 3.5 PN order, the SPA Fourier transform of  GW waveform of the $k$-th detector is given by~\cite{Sathyaprakash:2009xs}
\begin{align}
	h_k(f)=&\mathcal A_k f^{-7/6}{\rm exp}
	\{i[2\pi f t_{\rm c}-\pi/4-2\psi_c+2\Psi(f/2)]\nonumber\\ &-\varphi_{k,(2,0)}\},
\end{align}
where the Fourier amplitude $\mathcal A_k$ is given by
\begin{align}
	\mathcal A_k=&\frac{1}{d_{\rm L}}\sqrt{(F_{+,k}(1+\cos^{2}\iota))^{2}+(2F_{\times,k}\cos\iota)^{2}}\nonumber\\ &\times\sqrt{5\pi/96}\pi^{-7/6}\mathcal M^{5/6}_{\rm chirp},
\end{align}
where $d_{\rm L}$ is the luminosity distance to the source, $\mathcal M_{\rm chirp}$ is the chirp mass of the binary system, and the detailed forms of $\psi(f/2)$ and $\varphi_{k,(2,0)}$ could be found in Refs.~\cite{Zhao:2017cbb,Blanchet:2004bb}.
Under SPA, $F_{+,k}$, $F_{\times,k}$ and $\Phi_{ij}$ are all functions with respect to frequency, which are given by
\begin{equation}
\begin{aligned}
	&F_{+,k}(f)=F_{+,k}(t=t_{\rm f}), ~~~~F_{\times,k}(f)=F_{\times,k}(t=t_{\rm f}),\\
	&\Phi_{ij}(f)=\Phi_{ij}(t=t_{\rm f}),
\end{aligned}
\end{equation}
where $t_{\rm f}=t_{\rm c}-(5/256)\mathcal M^{-5/3}_{\rm chirp}(\pi f)^{-8/3}$ and $t_{\rm c} \in [0,10]$~yr  is the coalescence time~\cite{Maggiore:2007ulw}.

The term $t_{\rm f}$, which is mentioned above, represents the effect of the movement of the Earth during the time of the GW signal. If this effect is ignored, $t_{\rm f}$ can be approximately treated as a constant for a given GW event. For binary coalescence, the duration of the signal $t_*$ in a detector band is a strong function of the detector’s low-frequency cutoff $f_{\rm lower}$~\cite{Maggiore:2007ulw},
\begin{equation}
	t_*=0.86~{\rm day} \left(\frac{1.21~M_{\odot}}{\mathcal{M}_{\rm chirp}}\right)^{5/3} \left(\frac{2~{\rm Hz}}{f_{\rm lower}}\right)^{8/3}.
\end{equation}
For 3G GW detectors, $f_{\rm lower}$ is extended to about 1 Hz. For the BNS with $m_1=m_2=1.4~M_{\odot}$, we have $t_*=5.44~\rm {days}$. Therefore, the impact of the Earth’s rotation is important. For this reason, we consider this effect in our analysis.

Having the BNS coalescence catalog, we can easily determine whether its generated GW emission could be detected by GW detectors. Here we adopt the signal-to-noise ratio (SNR) threshold to be 12.\footnote{Here we approximate the GW detection with an unphysical cut on the true parameters of each event. Note that real searches are not equipped with access to the optimal SNR.} For low-mass systems, the combined SNR for the detection network of $N$ independent detectors is given by

\begin{equation}
\rho=(\tilde{\boldsymbol h}|\tilde{\boldsymbol h})^{1/2},       
\end{equation}
where $\tilde{\boldsymbol h}$ is the frequency-domain GW waveform of $N$ independent detectors as mentioned in Eq.~(\ref{eq:waveform}).
The inner product is defined as
\begin{equation}
(\boldsymbol a|\boldsymbol b)=2\int_{f_{\rm lower}}^{f_{\rm upper}}\{\boldsymbol a(f)\boldsymbol b(f)^*+\boldsymbol b(f)\boldsymbol a(f)^*\}{\rm d}f,
\end{equation}
where $\boldsymbol a$ and $\boldsymbol b$ are column matrices of the same dimension, * represents conjugate transpose, $f_{\rm lower}$ is the lower cutoff frequency ($f_{\rm lower}=1$ Hz for ET and $f_{\rm lower}=5$ Hz for CE), and $f_{\rm upper}=2/(6^{3/2}2\pi M_{\rm obs})$ is the frequency at the last stable orbit with $M_{\rm obs}=(m_1+m_2)(1+z)$. 

\subsection{Detection of short GRBs}

A structured GRB jet has an angular dependence on energy and bulk Lorentz factor, and is generally described by an ultra-relativistic core without sharp edges that smoothly transforms to a milder relativistic outflow at greater angles. Typical angular profiles are provided by Gaussian or power-law jet models. Given the uncertainty provided by one firm observation, the majority of late-time EM follow up campaigns have considered the former model~\cite{Resmi:2018wuc}. According to the observations of GW170817/GRB170817A, the jet profile model of short GRB is given by~\cite{Howell:2018nhu}

\begin{equation}
L_{\rm iso}(\theta_{\rm v})=L_{\rm on}{\rm exp}\left(-\frac{\theta^2_{\rm v}}{2\theta^2_{\rm c}} \right),
\label{eq:jet}
\end{equation}
where $L_{\rm iso}(\theta_{\rm v})$ is the isotropically equivalent luminosity of short GRB observed at different viewing angles $\theta_{\rm v}$, $L_{\rm on}$ is the on-axis isotropic luminosity defined by $L_{\rm on} \equiv L_{\rm iso}(0)$, and $\theta_{\rm c}$ is the characteristic angle of the core, which is given by $\theta_{\rm c}=4.7^{\circ}$. In this paper, we assume the directions of the jets are aligned with the binary orbital angular momentum, namely $\iota=\theta_{\rm v}$.

The detection probability of a short GRB is determined by $\Phi(L){\rm d}L$, where $\Phi(L)$ is the intrinsic luminosity function and $L$ is the peak luminosity of each burst assuming isotropic emission in the rest frame in the 1--10000~keV energy range. In this paper, we assume an empirical broken-power-law luminosity function when considering the short GRBs 

\begin{equation}
	\Phi(L)\propto
	\begin{cases}
		(L/L_*)^{\alpha_{\rm L}}, & L<L_*, \\
		(L/L_*)^{\beta_{\rm L}}, & L\ge L_*,
	\end{cases}
	\label{eq:distribution}
\end{equation}
where $L_{*}$ is characteristic luminosity separating the low and high end of the luminosity function, $\alpha_{\rm L}$ and $\beta_{\rm L}$ are the characteristic slopes describing these regimes, respectively. Following Ref.~\cite{Wanderman:2014eza}, we adopt $\alpha_{\rm L}=-1.95$, $\beta_{\rm L}=-3$, and $L_{*}=2\times10^{52}$ erg sec$^{-1}$. Here, we term the on-axis isotropic luminosity $L_{\rm on}$ as the peak luminosity $L$. We also assume a standard low end cutoff in luminosity of $L_{\rm min} = 10^{49}$ erg sec$^{-1}$.

To determine the detection probability of a short GRB by Eq.~(\ref{eq:distribution}), we need to convert the flux limit of the GRB satellite $P_{\rm T}$ to the isotropically equivalent luminosity $L_{\rm iso}$. For the 3G GW detectors, we assume here that the future THESEUS-like telescope with its X-$\gamma$ ray Imaging Spectrometer (XGIS) can make a coincident detection. A GRB detection is recorded if the value of the observed flux is greater than the flux limit $P_{\rm T}=0.2~\rm{ph~s^{-1}~cm^{-2}}$, in the 50-300 keV band for THESEUS telescope~\cite{Stratta:2018ldl}. We use the standard fiux-luminosity relation with two corrections: an energy normalisation

\begin{equation}
C_{\rm det}=\frac{\int^{10000~\rm{keV}}_{1~\rm{keV}}EN(E){\rm d}E}{\int^{E_{\rm max}}_{E_{\rm min}}N(E){\rm d}E},
\end{equation}
and a $k$-correction

\begin{equation}
k(z)=\frac{\int^{E_{\rm max}}_{E_{\rm min}}N(E){\rm d}E}{\int^{E_{\rm max}(1+z)}_{E_{\rm min}(1+z)}N(E){\rm d}E},
\end{equation}
where $[E_{\rm min},E_{\rm max}]$ is the detector's energy window~\cite{Howell:2018nhu,Wanderman:2014eza}. The observed photon flux is scaled by $C_{\rm det}$ to account for the missing fraction of the $\gamma$-ray energy seen in the detector band. The cosmological $k$-correction is due to the redshifted photon energy when traveling from source to detector. $N(E)$ is the observed GRB photon spectrum in units of $\rm{ph~s^{-1}~keV^{-1}~cm^{-2}}$.
For short GRB, the function $N(E)$ is simulated by the Band function~\cite{Band:2002te} which is  a function of spectral indices $(\alpha_{\rm B}, \beta_{\rm B})$ and break energy $E_{\rm b}$, 

\begin{widetext}
\begin{equation}
N(E)=\begin{cases}
N_0\left(\frac{E}{100~\rm keV}\right)^{\alpha_{\rm B}}\exp(-\frac{E}{E_0}),&E\leq E_{\rm b}, \\
 \\N_0\left(\frac{E_{\rm b}}{100~\rm keV}\right)^{\alpha_{\rm B}-\beta_{\rm B}}\exp(\beta_{\rm B}-\alpha_{\rm B}) \left(\frac{E}{100~\rm keV}\right)^{\beta_{\rm B}},&E>E_{\rm b},
\end{cases}
\end{equation}
\end{widetext}
here $E_{\rm b}=(\alpha_{\rm B}-\beta_{\rm B})E_0$ and $E_{\rm p}=(\alpha_{\rm B}+2)E_0$. From Ref.~\cite{Wanderman:2014eza}, we adopt $\alpha_{\rm B}=-0.5$, $\beta_{\rm B}=-2.25$, and a peak energy $E_{\rm p}=800~\rm keV$ in the source frame. This is a phenomenological fit to the observed spectra of GRB prompt emissions.
According to the relation between flux and luminosity for GRB~\cite{Meszaros:1995dj,Meszaros:2011zr}, we can convert the flux limit $P_{\rm T}$ to the luminosity by

\begin{equation}
L_{\rm iso}=4\pi d_{\rm L}^2(z)k(z)C_{\rm det}/(1+z)P_{\rm T}.
\end{equation}
Then, the value of the on-axis luminosity $L_{\rm on}$ can be given by Eq.~(\ref{eq:jet}).
Finally, using Eq.~(\ref{eq:distribution}), we can select the GRB detection  from the BNS samples by sampling $\Phi(L){\rm d}L$.

\subsection{Fisher information matrix and error analysis}

The Fisher information matrix of a GW detector network is given by

\begin{equation}
F_{ij}=\left(\frac{\partial \tilde{\boldsymbol{h}}}{\partial \theta_i}\Bigg |\frac{\partial \tilde{\boldsymbol{h}}}{\partial \theta_j}\right)
\end{equation}
where $\boldsymbol\theta$ denotes nine GW source parameters ($d_{\rm L}$, $\mathcal{M}_{\rm chirp}$, $\eta$, $\theta$, $\phi$, $\iota$, $t_{\rm c}$, $\psi_{\rm c}$, $\psi$) for a GW event. The covariance matrix is equal to the inverse of the Fisher matrix, i.e., ${\rm Cov}_{ij}=(F^{-1})_{ij}$. Thus, the instrumental error of GW parameter $\theta _i$ is $\Delta\theta_i=\sqrt{{\rm Cov}_{ii}}$.

For the cosmological parameter estimations, we treat the detectable GW events belonging to the GW-GRB joint observations as standard sirens for the cosmological analysis since the redshifts can be measured by the follow-up afterglow observations under the accurate location of associated GRBs. Although the Fisher information matrix can also be used for estimating cosmological parameters, in this paper we employ the Markov-chain Monte Carlo analysis for ease of combining with the actual CMB+BAO+SN (CBS) data, which is commonly used in the literature~\cite{Cai:2017aea,Cai:2016sby,Wang:2018lun,Zhang:2019ylr,Zhang:2019loq,Zhang:2018byx,Li:2019ajo,Jin:2020hmc,Zhang:2019ple,Jin:2023zhi,Jin:2022tdf,Jin:2021pcv,Jin:2023tou,Wu:2022dgy,Wang:2019tto,Wang:2021srv,Zhao:2019gyk,Jin:2023sfc,Hou:2022rvk,Li:2023gtu,Qi:2021iic}. We maximize the likelihood $\mathcal{L}\propto(-\chi^2/2)$ and infer the posterior probability distributions of cosmological parameters $\vec{\Omega}$. The $\chi^2$ function is defined as
\begin{align}
	\chi^2=\sum\limits_{i=1}^{N}\left[\frac{{d}_{\rm L}^i-d_{\rm L}({z}_i;\vec{\Omega})}{{\sigma}_{d_{\rm L}}^i}\right]^2,
\end{align}
where ${z}_i$, ${d}_{\rm L}^i$, and ${\sigma}_{d_{\rm L}}^i$ are the $i$-th GW event's redshift, luminosity distance, and the total error of the luminosity distance, respectively.

For the total error of the luminosity distance $d_{\rm L}$, we consider the instrumental error $\sigma_{d_{\mathrm{L}}}^{\mathrm{inst}}$ estimated by the Fisher information matrix, the weak-lensing error $\sigma_{d_{\mathrm{L}}}^{\mathrm{lens}}$, and the peculiar velocity error $\sigma_{d_{\mathrm{L}}}^{\mathrm{pv}}$~\cite{Jin:2021pcv,Wu:2022dgy,Jin:2022tdf,Jin:2023zhi,Jin:2023tou}. The total error of $d_{\rm L}$ is
\begin{equation}
	\left(\sigma_{d_{\mathrm{L}}}\right)^{2}=\left(\sigma_{d_{\mathrm{L}}}^{\mathrm{inst}}\right)^{2}+\left(\sigma_{d_{\mathrm{L}}}^{\text {lens }}\right)^{2}+\left(\sigma_{d_{\mathrm{L}}}^{\text {pv }}\right)^{2}.\label{eq:total}
\end{equation}
The error caused by weak lensing is given in Refs.~\cite{Hirata:2010ba,Tamanini:2016zlh,Speri:2020hwc},
\begin{align}
\sigma_{d_{\rm L}}^{\rm lens}(z)=&\left[1-\frac{0.3}{\pi/2} \arctan(z/0.073)\right]\times d_{\rm L}(z)\nonumber\\ &\times 0.066\left [\frac{1-(1+z)^{-0.25}}{0.25}\right ]^{1.8}.\label{lens}
\end{align}
The error caused by the peculiar velocity of the GW source is adopted from Ref.~\cite{Kocsis:2005vv} 

\begin{equation}
	\sigma_{d_{\rm L}}^{\rm pv}(z)=d_{\rm L}(z)\times \left [ 1+ \frac{c(1+z)^2}{H(z)d_{\rm L}(z)}\right ]\frac{\sqrt{\langle v^2\rangle}}{c},\label{pv}
\end{equation}
where $H(z)$ is the Hubble parameter and $c$ is the speed of light in vacuum. $\sqrt{\langle v^2\rangle}$ is the peculiar velocity of the GW source and we roughly set $\sqrt{\langle v^2\rangle}=500\ {\rm km\ s^{-1}}$, in agreement with the average value of the galaxy catalogs \cite{He:2019dhl}.

To date, the vast majority of redshift determinations of GRBs depend on optical to Near Infra-Red afterglow spectra (that unambiguously pinpoints the host galaxy) obtained from ground-based follow-up observations. Given that spectroscopic measurement renders the error from the redshift measurement of the EM counterpart negligible, we ignore the redshift measurement error of the GRB in this paper~\cite{THESEUS:2021uox}.

\section{Observation of GWs and GRBs}\label{sec3}

\begin{figure}[htbp]
	\includegraphics[width=0.9\linewidth,angle=0]{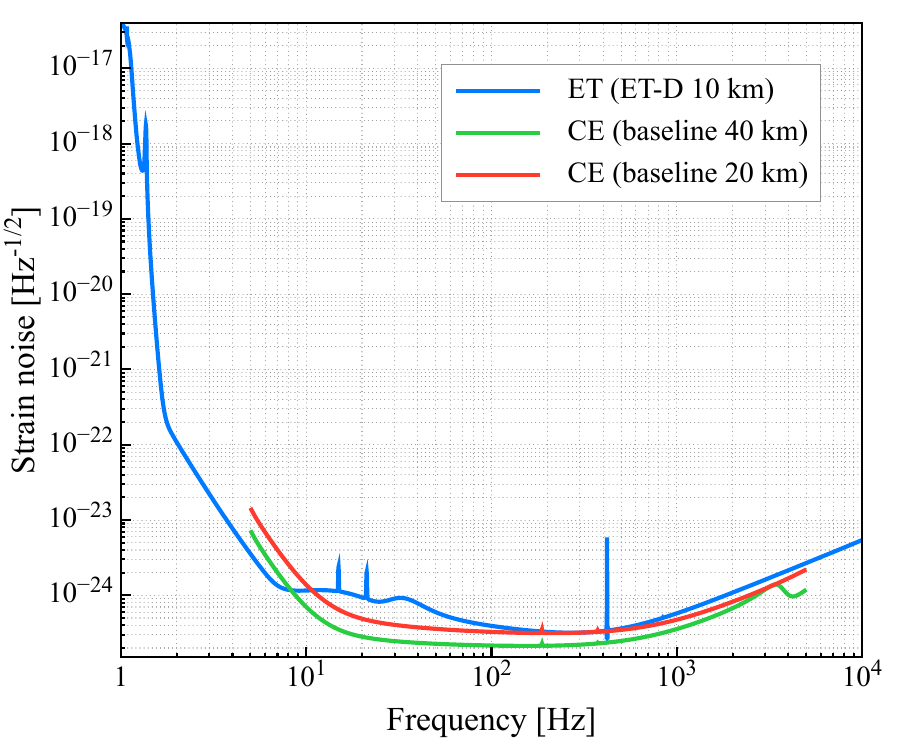}
	\caption{\label{fig1} Sensitivity curves of the 3G GW detectors considered in this work.}
\end{figure}

 In this section, we will discuss the BNS mergers’ detection rates and distributions of GW observations using 3G GW detectors, short GRB observations using a $\gamma$-ray detector with the characteristic of THESEUS, and the joint GW-GRB observations. We consider four different cases of 3G GW observations, single ET, single CE, the CE-CE network (two CE-like detectors, one in the US with 40 km arm length and another one in Australia with 20 km arm length, abbreviated as 2CE hereafter), and the ET-CE-CE network (one ET detector and two CE-like detectors, abbreviated as ET2CE hereafter). We adopt the sensitivity curve of ET from Ref.~\cite{ETcurve-web} and the sensitivity curves of CE from Ref.~\cite{CEcurve-web}, as shown in Fig.~\ref{fig1}. For the GW detector, in view of the high uncertainty of the duty cycle, we only calculate the best case where each detector has a duty cycle of 100\%~\cite{Zhu:2021ram}. The specific parameters characterizing the GW detector geometry
(latitude $\varphi$, longitude $\lambda$, opening angle $\zeta$, and arm bisector angle $\gamma$) are listed in Table~\ref{tab1}.

\begin{table*}[!htb]
	\caption{The specific coordinate parameters considered in this work.}
	\label{tab1} 
	\setlength\tabcolsep{20pt}
	\renewcommand{\arraystretch}{1.5}
	\begin{tabular}{ccccc}
		\hline \hline
		GW detector  & $\varphi\ (\mathrm{deg})$ & $\lambda\ (\mathrm{deg})$ & $\gamma\ (\mathrm{deg})$ & $\zeta\ (\mathrm{deg})$ \\
		\hline
		Einstein Telescope, Europe &  $40.443$ & $9.457$ & $0.000$ & 60 \\
		Cosmic Explorer, USA &  $43.827$ & $-112.825$ & $45.000$ & 90 \\
		Cosmic Explorer, Australia &  $-34.000$ & $145.000$ & $90.000$ & 90 \\
		\hline\hline
	\end{tabular}
\end{table*}

For the THESEUS-like telescope, we make the assumption of an 80\% duty cycle, which takes into account a 20\% reduction when the satellite passes through the Southern Atlantic Anomaly. Meanwhile, we consider a sky coverage fraction of 0.5~\cite{Stratta:2018ldl}. According to the THESEUS papers~\cite{THESEUS:2017qvx,THESEUS:2017wvz,Stratta:2018ldl}, the XGIS of THESEUS will be able to localize the source to around 5 arcmin, only within the central 2 sr of its field of view (FOV). If outside this central region, localization will be coarse at best. In this paper, we consider two scenarios. The first scenario, termed ``optimistic," assumes that all short GRBs detected by XGIS can provide perfect redshift estimates through follow-up observations. The second scenario, termed ``realistic," assumes that only about one-third of the short GRBs can provide perfect redshift estimates through follow-up observations \cite{Belgacem:2019tbw}.


Due to limitations in FOV and sampling of luminosity function, the results of each calculation of GW-GRB detection are slightly different. In order to obtain robust analysis results, we perform the calculation of the number of GW-GRB 50 times for the following discussion.

\begin{figure}[htbp]
	\includegraphics[width=0.9\linewidth,angle=0]{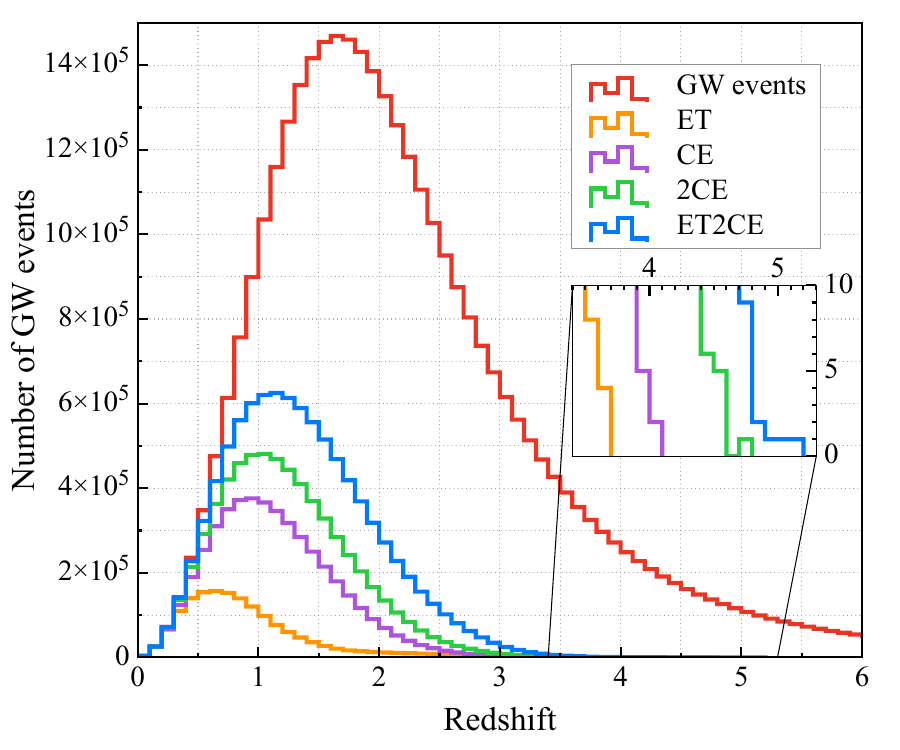}
	\caption{\label{fig2} Redshift distributions of the total GW events and the GW events detected by ET, CE, 2CE, and ET2CE assuming a 10-year observation.}
\end{figure}

\begin{figure}[htbp]
	\includegraphics[width=0.9\linewidth,angle=0]{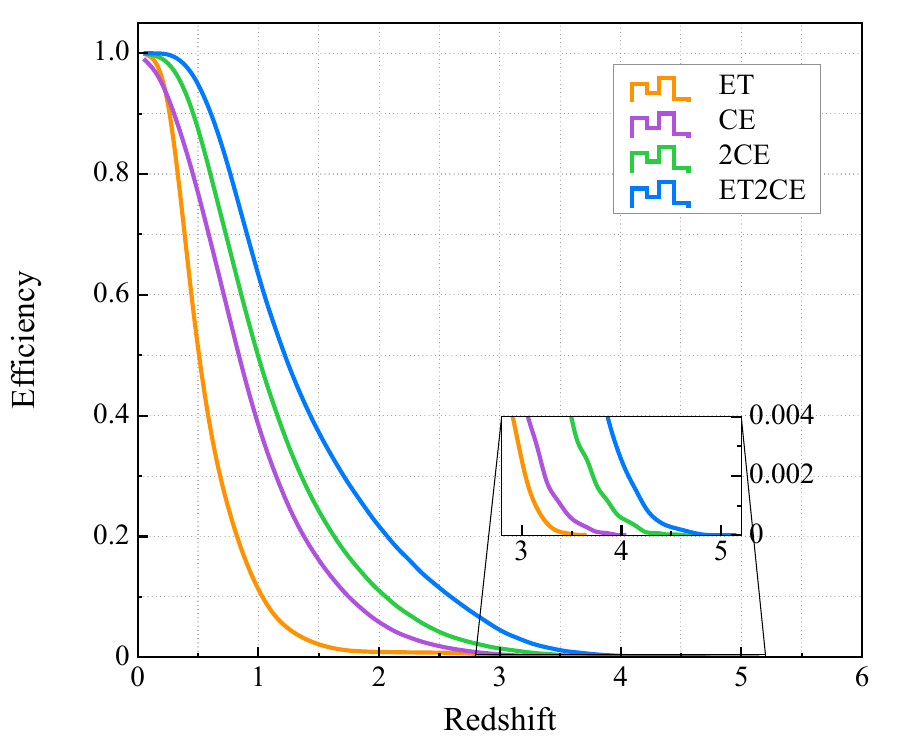}
	\caption{\label{fig3} Detection efficiencies of the 3G GW detectors.}
\end{figure}

We first calculate the redshift distributions of the total GW events and the GW events detected by ET, CE, the 2CE network, and the ET2CE network assuming a 10-year observation, as shown in Fig.~\ref{fig2}. We see that the number of GW events detected by CE is around three times that of ET. Meanwhile, the number of GW events detected by the GW detector network is significantly more than that of the single GW observatory. In Fig.~\ref{fig3}, we also show the detection efficiencies of GW detectors for comparison. For ET and CE alone, the horizon redshifts are up to about 3.8 and 4.1, with 50\% detection efficiencies at $z = 0.5$ and 0.8. For the 2CE and ET2CE network, the horizon redshifts extend to about 4.8 and 5.2, with 50\% detection efficiencies at $z = 1$ and 1.2, respectively.

\begin{figure}[htbp]
	\includegraphics[width=0.9\linewidth,angle=0]{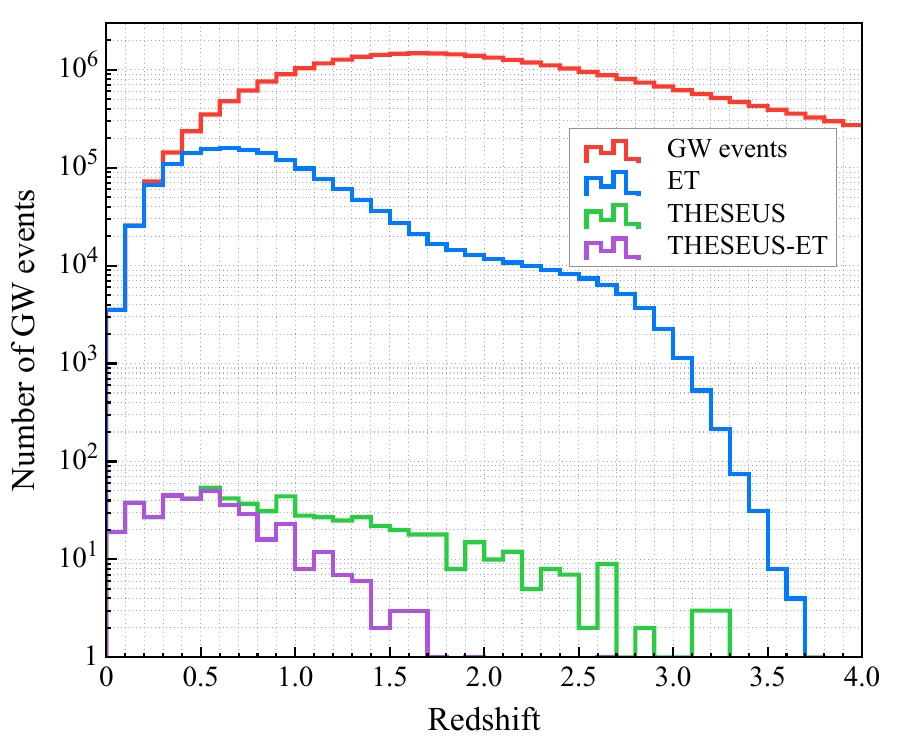}
	\caption{\label{fig4} Redshift distributions of the total GW events, the events detected by ET, the events detected by THESEUS, and the joint detections of ET+THESEUS assuming a 10-year observation.}
\end{figure}

Then, we select the GW events that can be triggered by both GW detectors and a THESEUS-like telescope. Fig.~\ref{fig4} shows the redshift distributions of the joint detection, in the specific case of THESEUS in synergy with ET. We can see that the redshift distribution of the GW-GRB detections mainly depends on the flux-limited GRB instrument (in the case of THESEUS). Only a very small fraction of the GW events can be finally identified by the short GRB detector. 

\begin{figure}[htbp]
	\includegraphics[width=0.9\linewidth,angle=0]{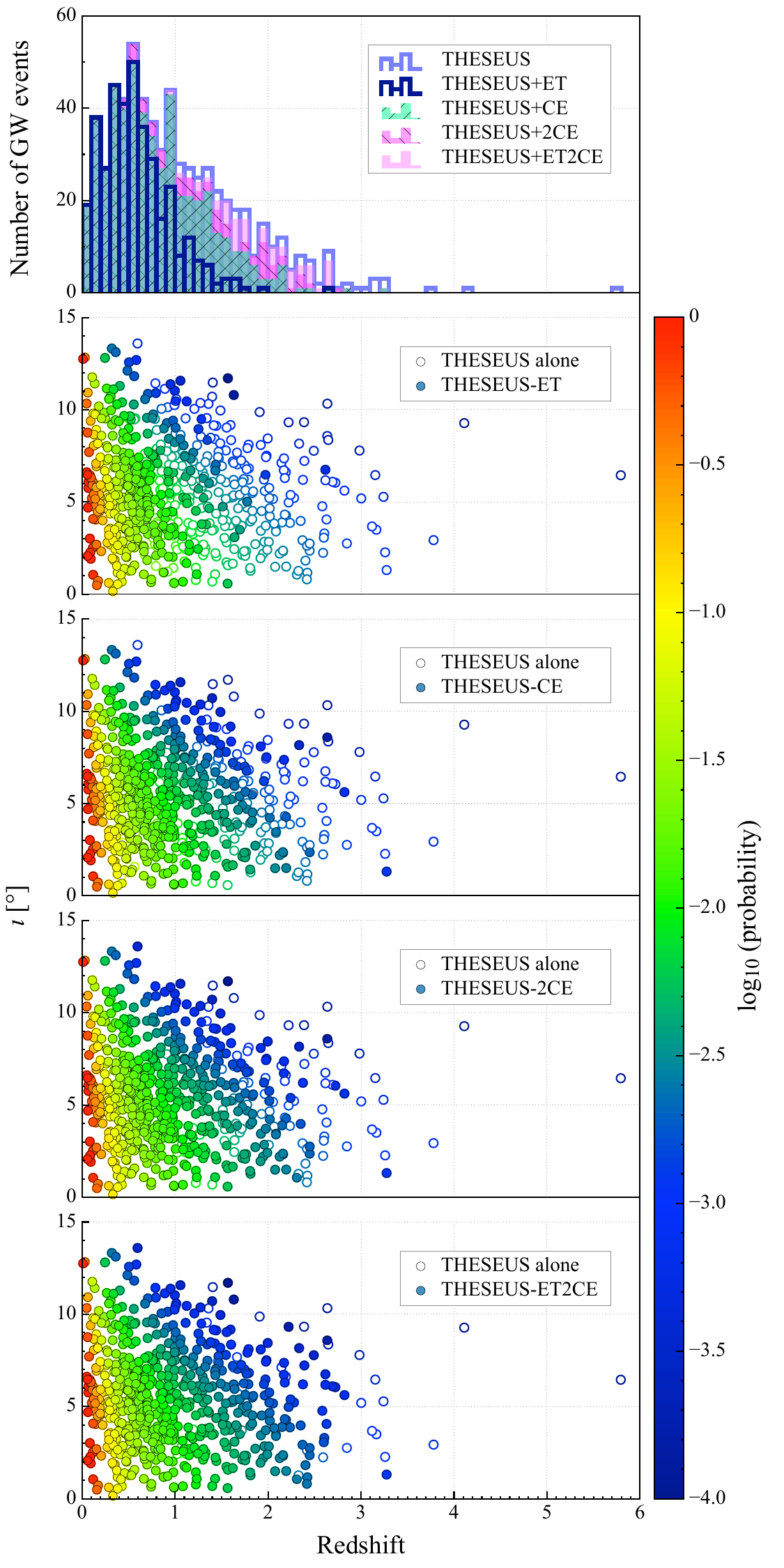}
	\caption{\label{fig5} Redshift distributions of short GRBs and GW-GRB coincidences for a 10-year observation in optimistic scenario for the FOV. {Top panel}: Redshift distributions of BNS detected by THESEUS and THESEUS in synergy with ET, CE, 2CE, and ET2CE. {Lower four panels}: The distributions of inclination angles $\iota$ and the redshifts of BNS samples, which can be triggered by THESEUS alone and THESEUS in synergy with ET, CE, 2CE, and ET2CE, respectively. The color bar indicates the logarithm of detection probability for THESEUS.}
\end{figure}

\begin{figure}[htbp]
	\includegraphics[width=0.9\linewidth,angle=0]{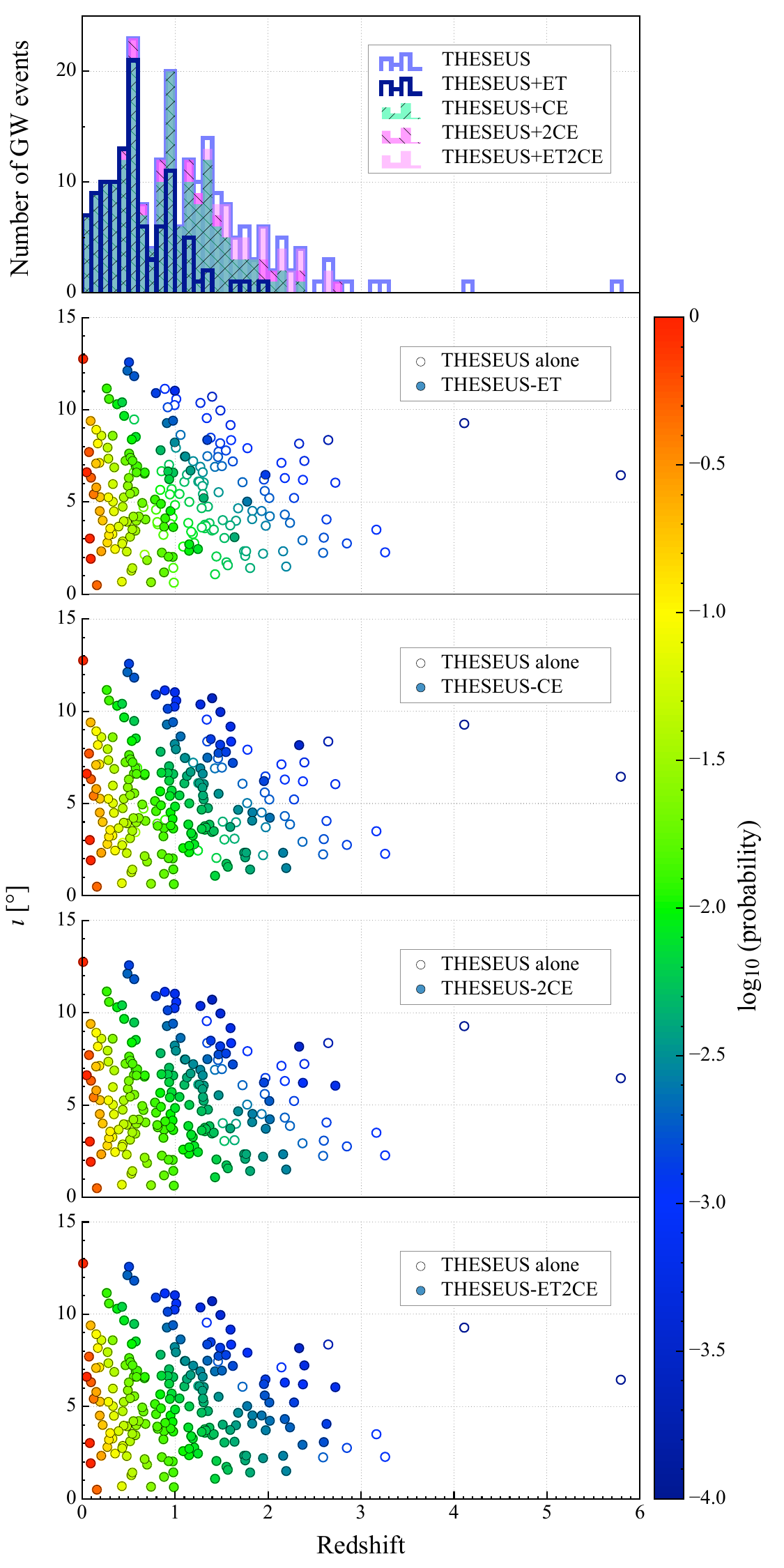}
	\caption{\label{fig6} Same as Fig.~\ref{fig5}, but assuming the realistic scenario for the FOV.}
\end{figure}

In Figs.~\ref{fig5} and~\ref{fig6}, we show the redshift distributions of short GRBs and GW-GRB coincidences for a 10-year observation in optimistic and realistic scenario for the FOV. The top panel shows the redshift distributions of BNS detections by the THESEUS and the THESEUS in synergy with ET, CE, the 2CE, and the ET2CE, respectively. For the ET alone in the optimistic and realistic scenarios, 52.5--59.1\% and 47.2--63.6\% of all the GRBs will have a detectable GW counterpart, respectively. For the CE alone in the two scenarios above, the proportions are 75.8--81.9\% and 75.6--85.2\%, which are about 20\% higher than ET alone. For a network of 3G GW observatories in the two scenarios above, we find that the vast majority of short GRBs detected in $\gamma$-ray have detectable GW counterparts, and the joint detection efficiencys approach about 90\%.  The lower four panels show the distributions of the inclination angles and the redshifts of BNS samples, which can be triggered by the THESEUS alone and the THESEUS synergy with ET, CE, 2CE, and ET2CE, respectively. The color bar indicates the logarithm of detection probability for THESEUS. We can see that due to the limitation of the Gaussian jet profile, the GW events that could be triggered by GW detectors and the GRB detector have inclination angles $\iota<15^{\circ}$. With the increase of redshift and inclination angle, the probability of the GRB detection decreases significantly. Under a structured Gaussian jet scenario, only emissions near the jet axis will be detected in $\gamma$-ray. Despite the fact that the ET2CE network detects a number of sources larger by an order of magnitude compared to the single ET (and to much larger redshift, see Fig.~\ref{fig2}), the corresponding joint GW-GRB detections do not follow the same increase, because of intrinsic limitations in the GRB detections.

\begin{table*}
	\caption{Numbers of BNS events detected by ET, CE, 2CE, ET2CE, and GRB events detected by THESEUS in a 10-year observation and the joint GW-GRB events triggered by THESEUS in synergy with ET, CE, 2CE, and ET2CE, respectively. Note that numbers in parenthesis show the number of sources with arcmin localization.}
	\label{tab2}
	\centering
    \setlength{\tabcolsep}{2.5mm}
    \renewcommand{\arraystretch}{2}
	\begin{tabular}{|c|c|c|c|c|c|c|c|} \hline
		BNS samples&Detection strategy&GW detections	&GRB detections	&GW-GRB detections\\ \hline
		\multirow{4}{*}{34430020}&      ET       &   1553981    &     \multirow{4}{*}{580-720 (193-240)}    &  334-412 (100-138)    \\ 
		\multirow{4}{*}{}                  &       CE       &   4645716    &     \multirow{4}{*}{}                                   &  463-575 (153-199)    \\  
		\multirow{4}{*}{}                  &      2CE      &  6434200   &     \multirow{4}{*}{}                                    &    514-635 (168-216)   \\  
		\multirow{4}{*}{}                  &    ET2CE   &   9342889 &     \multirow{4}{*}{}                                    &      558-674 (181-227)  \\  \hline
	\end{tabular}
\end{table*}

In Table~\ref{tab2}, we show the results of our simulations for the 3G era in terms of the number of GWs and short GRB signals from BNS mergers, along with the number of joint GW-GRB detections assuming a 10-year observation. Here the number of events with arcmin localization is shown in parenthesis. For the single ET detector, our estimate of GW detections are about $1.55\times 10^6$ in 10 years, consistent with the estimate in Ref.~\cite{Sathyaprakash:2009xt}. For GRB detections, we estimate that 580--720 short GRBs could be triggered by THESEUS, consistent with the order of magnitude in Ref.~\cite{THESEUS:2017wvz}. Because the assumed luminosity function and BNS merger rate differ from that assumed in Ref.~\cite{THESEUS:2017wvz}, the calculated number in this work is slightly higher than that in Ref.~\cite{THESEUS:2017wvz}. As can be seen from Table~\ref{tab2}, with the single ET detector, we should expect around 334--412 coincident GW-GRB events in 10-year observation, consistent with the order of magnitude in previous work~\cite{Hou:2022rvk,Chen:2020zoq,Yu:2021nvx}.

\section{Impact of the Earth's rotation}\label{sec4}

In this section, we focus on the impact of the Earth’s rotation when considering the luminosity distance uncertainties of BNS mergers for ET and ET2CE. To better show the comparison, we consider two scenarios, one accounting for the Earth's rotation and the other without. Note that the analysis is based on $10^4$ random BNS simulations at $z=0.5$.

\begin{figure*}[htbp]
	\includegraphics[width=0.7\linewidth,angle=0]{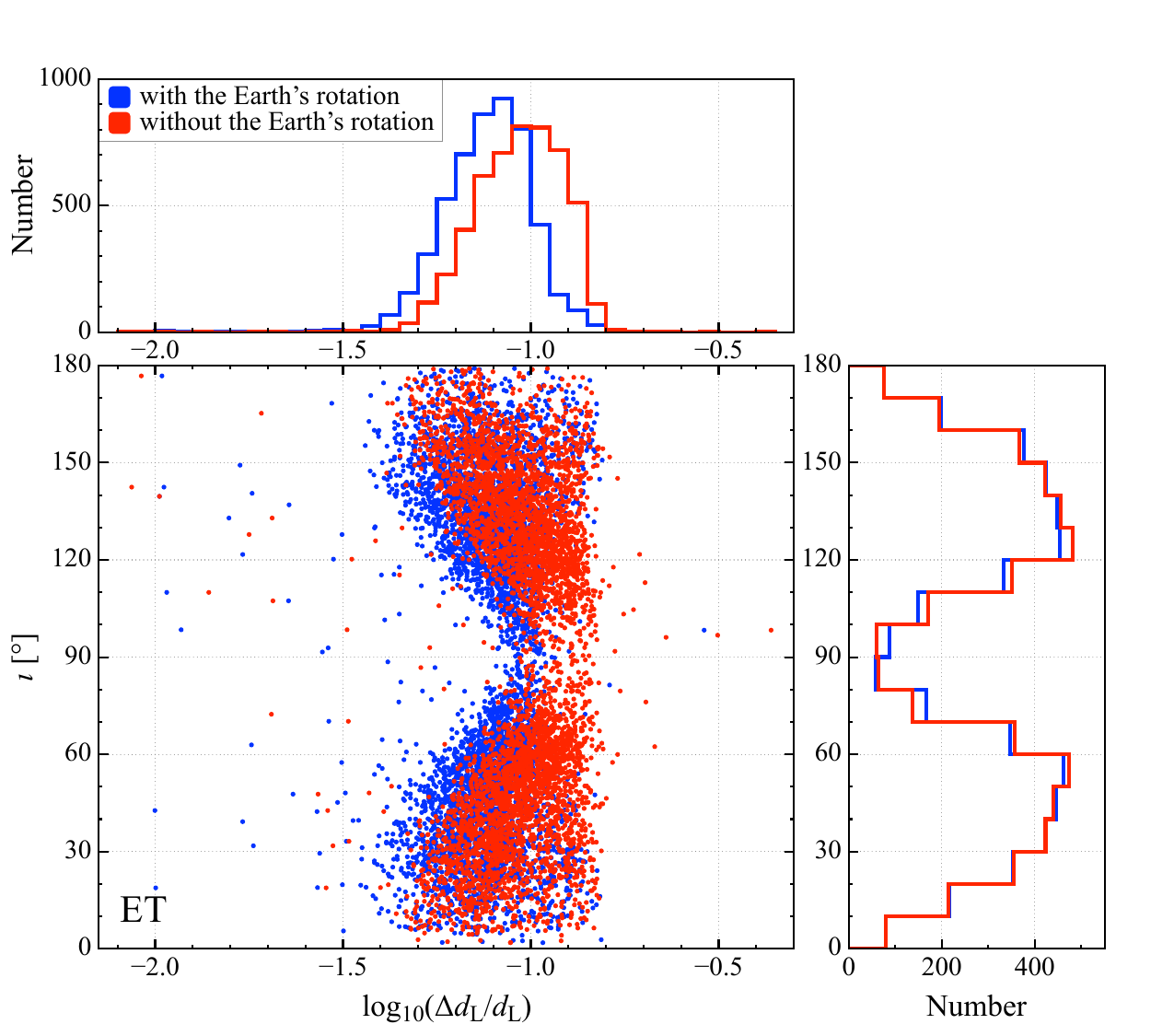}
	\includegraphics[width=0.7\linewidth,angle=0]{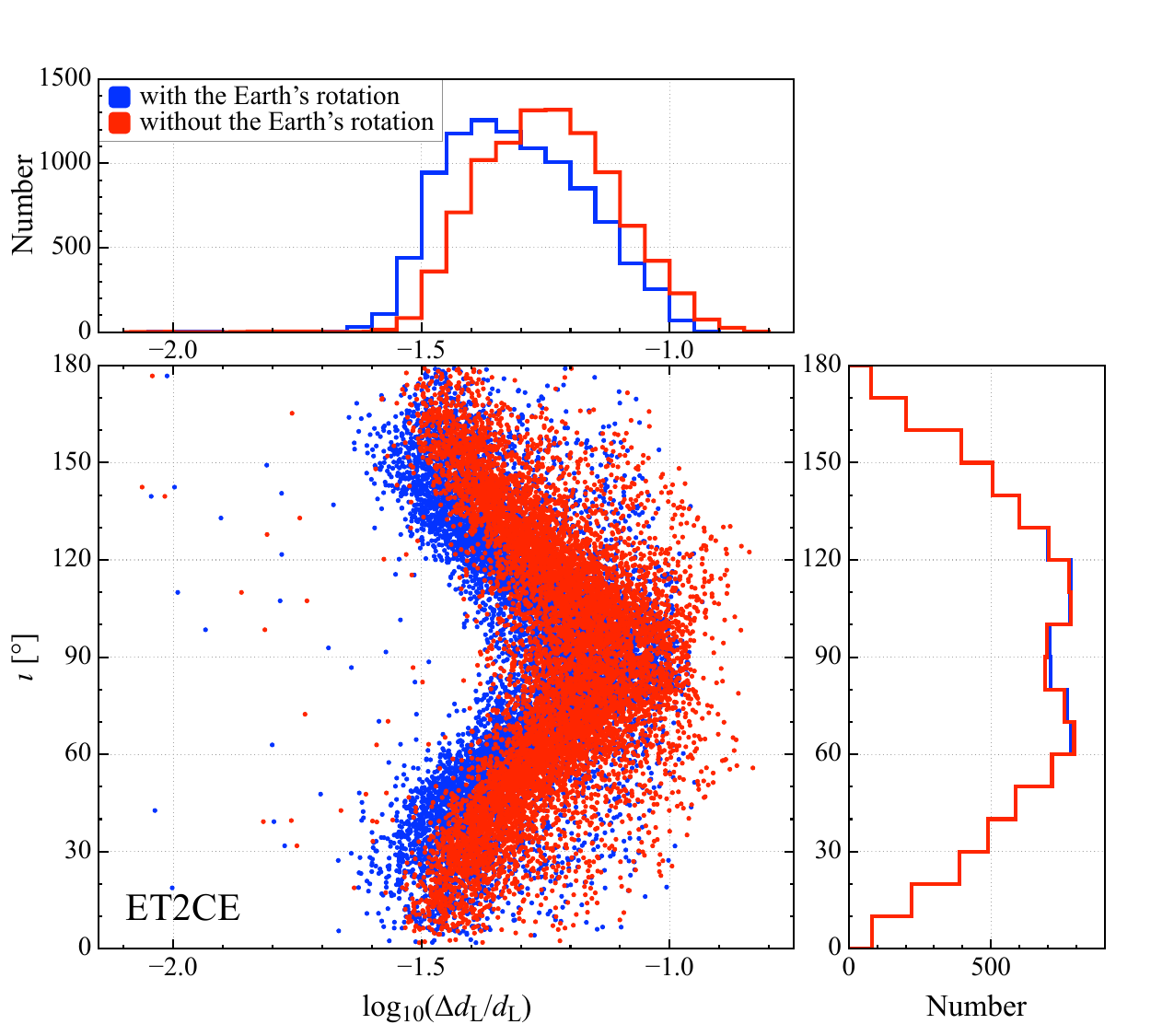}
	\caption{\label{fig7} Distributions of $\Delta d_{\rm L}/d_{\rm L}$ with respective to the inclination angle $\iota$ for $10^4$ BNS samples at the redshift $z=0.5$. The upper panel shows the results of ET and the lower panel shows those of ET2CE. In each panel, the red dots and histograms indicate the distribution without the Earth’s rotation, and the blue dots and histograms indicate the distribution with the Earth’s rotation.}
\end{figure*}

The effect of the Earth’s rotation for ET and ET2CE can be found in Fig.~\ref{fig7}. We can find that the luminosity distance uncertainties considering the Earth's rotation are less than those without considering the Earth’s rotation. Note that these results are obtained from the GW observations alone. This is primarily influenced by the following two aspects: One is the modulation of the Doppler effect quantified by the time-dependent function $\Phi_{ij}$, and the other is quantified by the time-dependent detector responses $F_{+,k}$ and $F_{\times,k}$. In this case, an individual detector can be effectively treated as a detector network with long baselines formed by the trajectory of the detector as it rotates with the Earth. Therefore, we consider the effect of the Earth’s rotation in the following cosmological analysis.

\section{Constraint results}\label{sec5}

In this section, we shall report the constraint results of cosmological parameters. We consider the $\Lambda$CDM, $w$CDM, $w_0w_a$CDM models, and IDE models (I$\Lambda$CDM and I$w$CDM) to complete the cosmological analysis. For $\Lambda$CDM, $w$CDM, and  $w_0w_a$CDM models, we constrain these cosmological models with GW standard siren data alone. Note that GW provides rather poor constraints on the IDE models and thus we only show the constraint results of CBS and CBS+GW. Meanwhile, we also give the constraint results of CBS and CBS+GW for all five cosmological models above to show the capability of GW standard sirens of breaking the cosmological parameter degeneracies. For the CMB data, we employ the $Planck$ TT, TE, EE spectra at $\ell\geq 30$, the low-$\ell$ temperature Commander likelihood, and the low-$\ell$ SimAll EE likelihood from the $Planck$ 2018 data release \cite{Planck:2018vyg}. For the BAO data, we adopt the measurements from 6dFGS ($z_{\rm eff}=0.106$) \cite{Beutler:2011hx}, SDSS-MGS ($z_{\rm eff}=0.15$) \cite{Ross:2014qpa}, and BOSS DR12 ($z_{\rm eff}=0.38$, 0.51, and 0.61) \cite{BOSS:2016wmc}. For the SN data, we employ the Pantheon sample consisting of 1048 data points \cite{Pan-STARRS1:2017jku}. The 1$\sigma$ and 2$\sigma$ posterior distribution contours for the cosmological parameters of interest are shown in Figs.~\ref{fig10}--\ref{fig14} and the 1$\sigma$ errors for the marginalized parameter constraints are shown in Tables~\ref{tab4}--\ref{tab7}. We use $\sigma(\xi)$ and $\varepsilon(\xi)$ to represent the absolute and relative errors of parameter $\xi$, with $\varepsilon(\xi)$ defined as $\varepsilon(\xi)=\sigma(\xi)/\xi$. Note that in the following, we take ET2CE as the representative of GW to make some relevant discussions.

In IDE models, there is a problem of early-time perturbation instability~\cite{Majerotto:2009zz,Clemson:2011an,He:2008si}, because the cosmological perturbations of dark energy in the IDE models will be divergent in a part of the parameter space, ruining the IDE cosmology in the perturbation level. To overcome the problem, Li \emph{et al.}~\cite{Li:2014eha,Li:2014cee,Li:2015vla,Li:2023fdk} established an effective theoretical framework for IDE cosmology based on the extended version of the PPF approach \cite{Fang:2008sn,Hu:2008zd} to the IDE models, referred to as the ePPF approach. The approach can safely calculate the cosmological perturbations in the whole parameter space in the IDE models. In our analysis of this paper, we employ the ePPF method to treat the cosmological perturbations (see e.g., Refs.~\cite{Zhang:2017ize,Feng:2018yew} for more details about the ePPF method).

The estimated numbers of standard sirens are crucial for the cosmological parameter estimations. However, the results of each calculation are slightly different. For the purpose of obtaining robust analysis results, we choose the estimate of joint GW-GRB detections corresponding to the median of GRBs in Sec.~\ref{sec3} as the final estimation, and treat them as standard sirens in the following cosmological analysis. In order to show the potential of GW standard sirens in estimating cosmological parameters, we consider the optimistic and realistic scenarios for FOV to make the cosmological analysis. The numbers of GW standard sirens in the following cosmological analysis are shown in Table~\ref{tab3}. And the redshift distributions are shown in Figs.~\ref{fig5} and~\ref{fig6}.

\begin{table*}
	\caption{Numbers of GW standard sirens in cosmological analysis, triggered by THESEUS assuming the optimistic and realistic scenarios in synergy with ET, CE, 2CE, and ET2CE, respectively.}
		\label{tab3} 
		\centering
		\setlength{\tabcolsep}{2.5mm}
		\renewcommand{\arraystretch}{2}
		\begin{tabular}{|c|*{4}{>{\centering\arraybackslash}m{1.75cm}|}}\hline
			Detection strategy&ET&CE&2CE&ET2CE\\ \hline
			Optimistic scenario&      368      &   512   &    571  &  621  \\ 
			Realistic scenario   &      107      &    171   &    186   &  206 \\   \hline
		\end{tabular}
\end{table*}

\begin{figure}[htbp]
	\includegraphics[width=0.9\linewidth,angle=0]{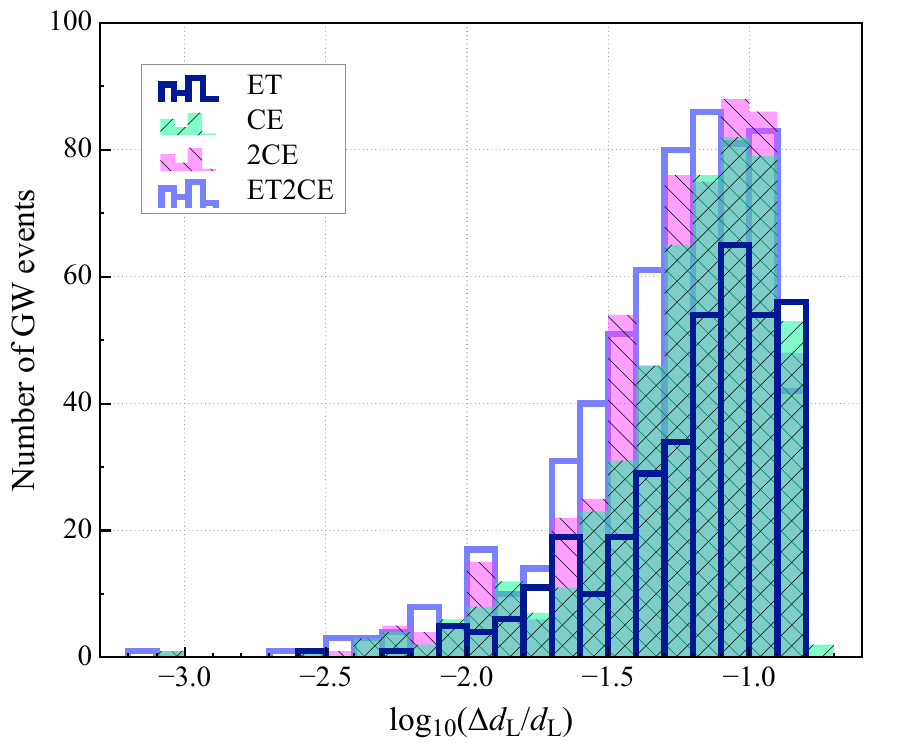}
	\caption{\label{fig8} Distributions of luminosity distance uncertainty $\Delta d_{\rm L}/d_{\rm L}$ of GW standard sirens for ET, CE, 2CE, and ET2CE in the optimistic scenario.}
\end{figure}

\begin{figure}[htbp]
	\includegraphics[width=0.9\linewidth,angle=0]{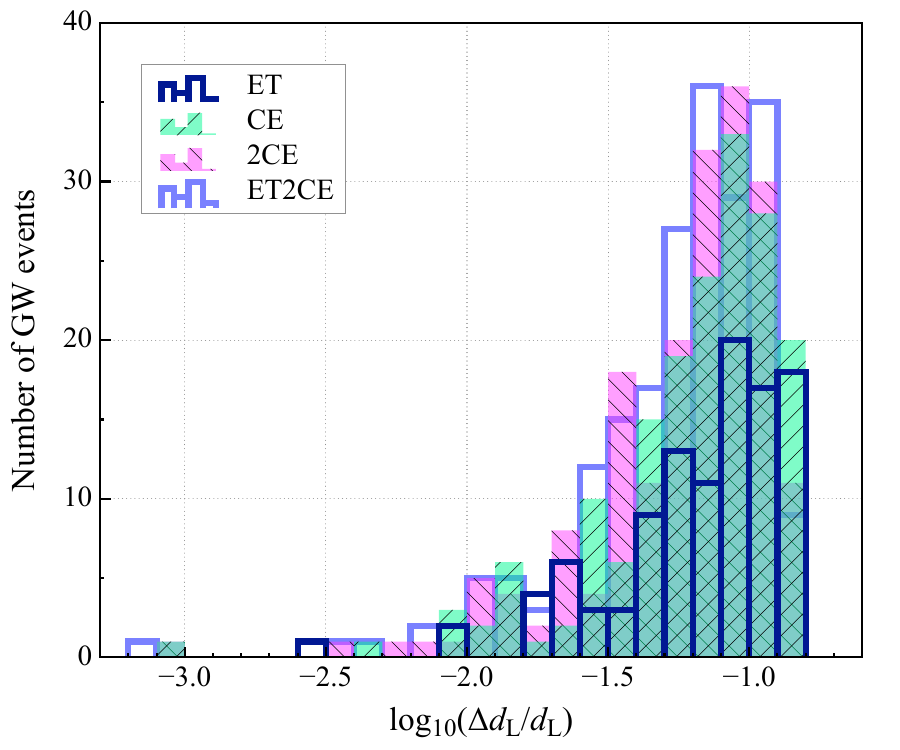}
	\caption{\label{fig9} Same as Fig.~\ref{fig8}, but assuming the realistic scenario.}
\end{figure}

In Figs.~\ref{fig8} and \ref{fig9}, we present the distributions of luminosity distance uncertainty $\Delta d_{\rm L}/d_{\rm L}$ of GW standard sirens for ET, CE, 2CE, and ET2CE in optimistic and realistic scenarios. We can see that the measurement precisions of luminosity distances are mainly 4\%--12\%. ET2CE gives the best measurement precisions of $d_{\rm L}$, followed by 2CE, CE, and ET.

\subsection{Constraint results in optimistic scenario}

\begin{figure*}[htbp]
	\includegraphics[width=0.45\linewidth,angle=0]{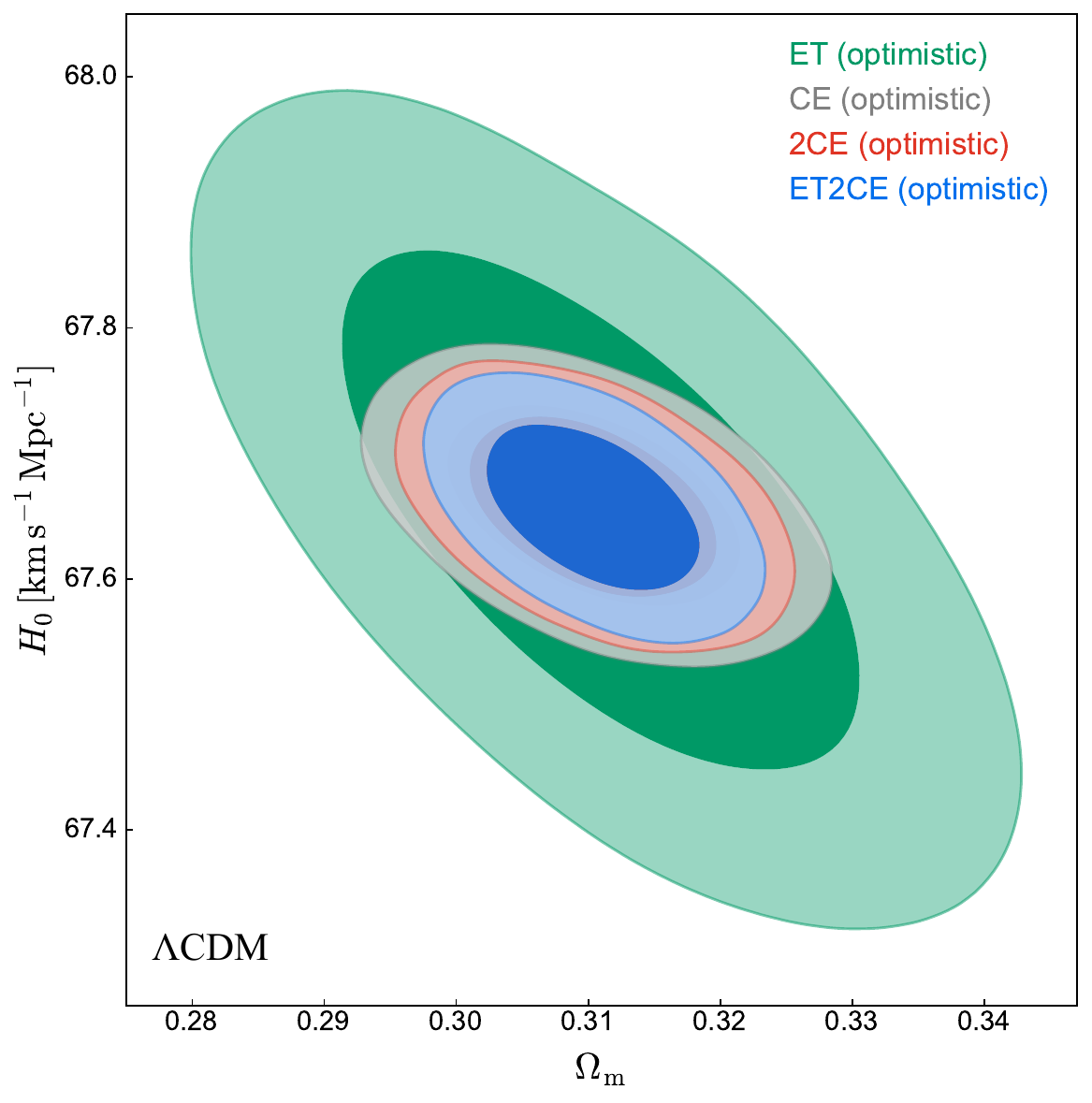}
    \includegraphics[width=0.45\linewidth,angle=0]{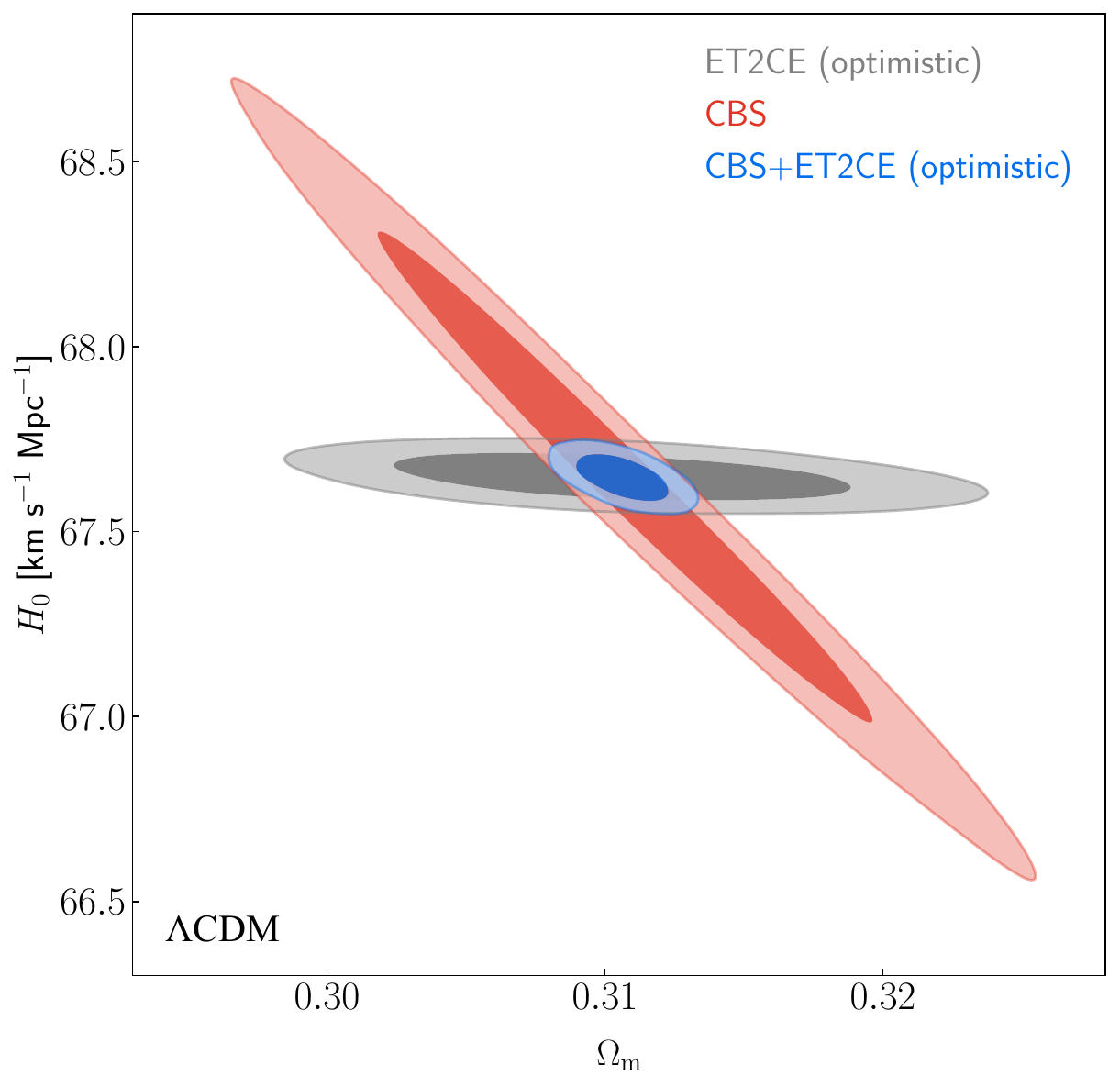}
	\caption{\label{fig10} Constraints on the $\Lambda$CDM model in the optimistic scenario. Left panel: Two-dimensional marginalized contours (68.3\% and 95.4\% confidence level) in the $\Omega_{\rm m}$--$H_0$ plane using the mock standard siren data of ET, CE, 2CE, and ET2CE. Right panel: Two-dimensional marginalized contours (68.3\% and 95.4\% confidence level) in the $\Omega_{\rm m}$--$H_0$ plane using the ET2CE, CBS, and CBS+ET2CE data.}
\end{figure*}

In Fig.~\ref{fig10}, we show the constraint results in the $\Omega_{\rm m}$--$H_0$ for the $\Lambda$CDM model. As can be seen, ET gives the worst constraint results, followed by CE, 2CE, and ET2CE. The prime cause is that the constraint results heavily depend on the numbers and the errors of the standard siren data, while ET2CE has the most number and the minimum error of standard siren data points, followed by 2CE, CE, and ET, as shown in Figs.~\ref{fig5} and~\ref{fig8}. Even so, ET gives $\sigma(H_0)=0.140~\rm km~s^{-1}~Mpc^{-1}$ with a constraint precision of 0.207\%, which is much better than that of CBS. However, ET gives a loose constraint on $\Omega_{\rm m}$ with a precision of 4.18\%, which is worse than that of CBS. When using the ET2CE data, the constraint results of $\Omega_{\rm m}$ and $H_0$ are both better than those of CBS. Meanwhile, we could clearly see that the parameter degeneracy orientations of ET2CE and CBS in the $\Omega_{\rm m}$--$H_0$ plane are different and thus the combination of them could break cosmological parameter degeneracies. With the addition of ET2CE to CBS, the constraint precisions of cosmological parameters are greatly improved. CBS+ET2CE gives $\sigma(\Omega_{\rm m})=0.0011$ and $\sigma(H_0)=0.041~\rm km~s^{-1}~Mpc^{-1}$, which are 81.4\% and 90.7\% better than those of CBS. Moreover, the constraint precisions of $\Omega_{\rm m}$ and $H_0$ are 0.35\% and 0.061\%, which are both much better than 1\%, the standard of precision cosmology.

\begin{figure*}[htbp]
    \includegraphics[width=0.45\linewidth,angle=0]{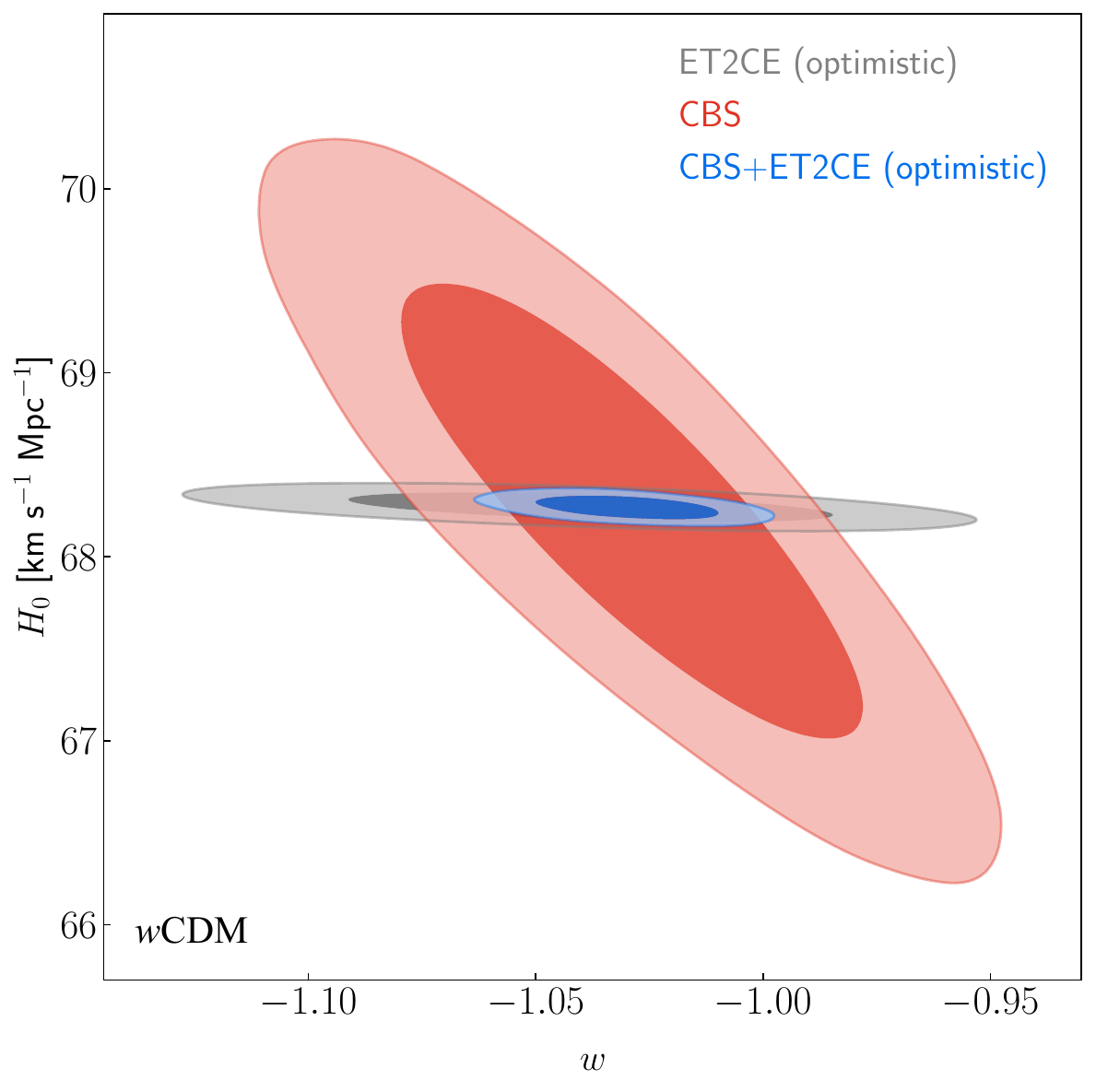}
    \includegraphics[width=0.45\linewidth,angle=0]{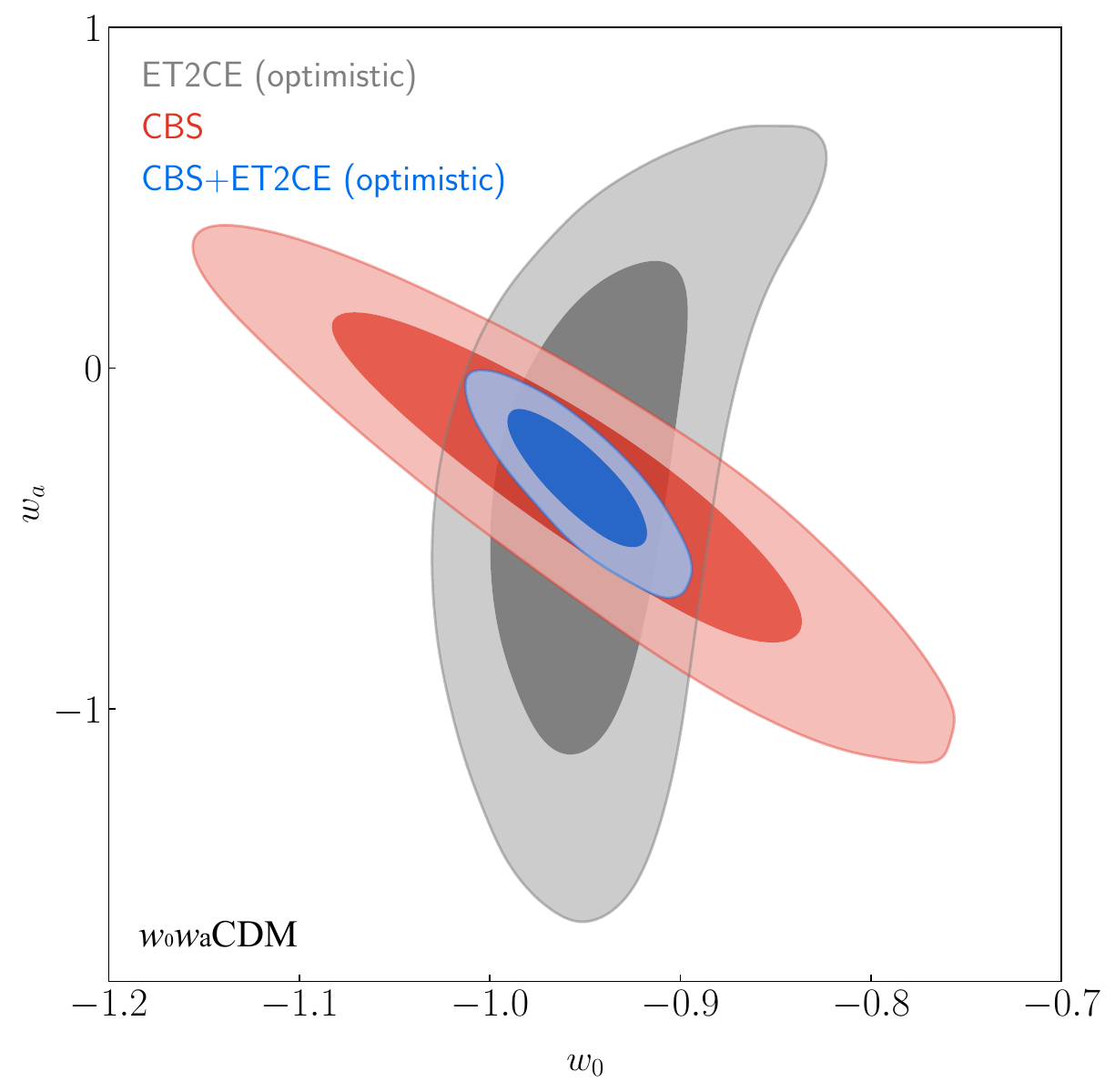}
	\caption{\label{fig11} Two-dimensional marginalized contours (68.3\% and 95.4\% confidence level) in the $w$--$H_0$ and $w_0$--$w_a$ planes using the ET2CE, CBS, and CBS+ET2CE data in the optimistic scenario.}
\end{figure*}

\begin{figure*}[htbp]
	\includegraphics[width=0.45\linewidth,angle=0]{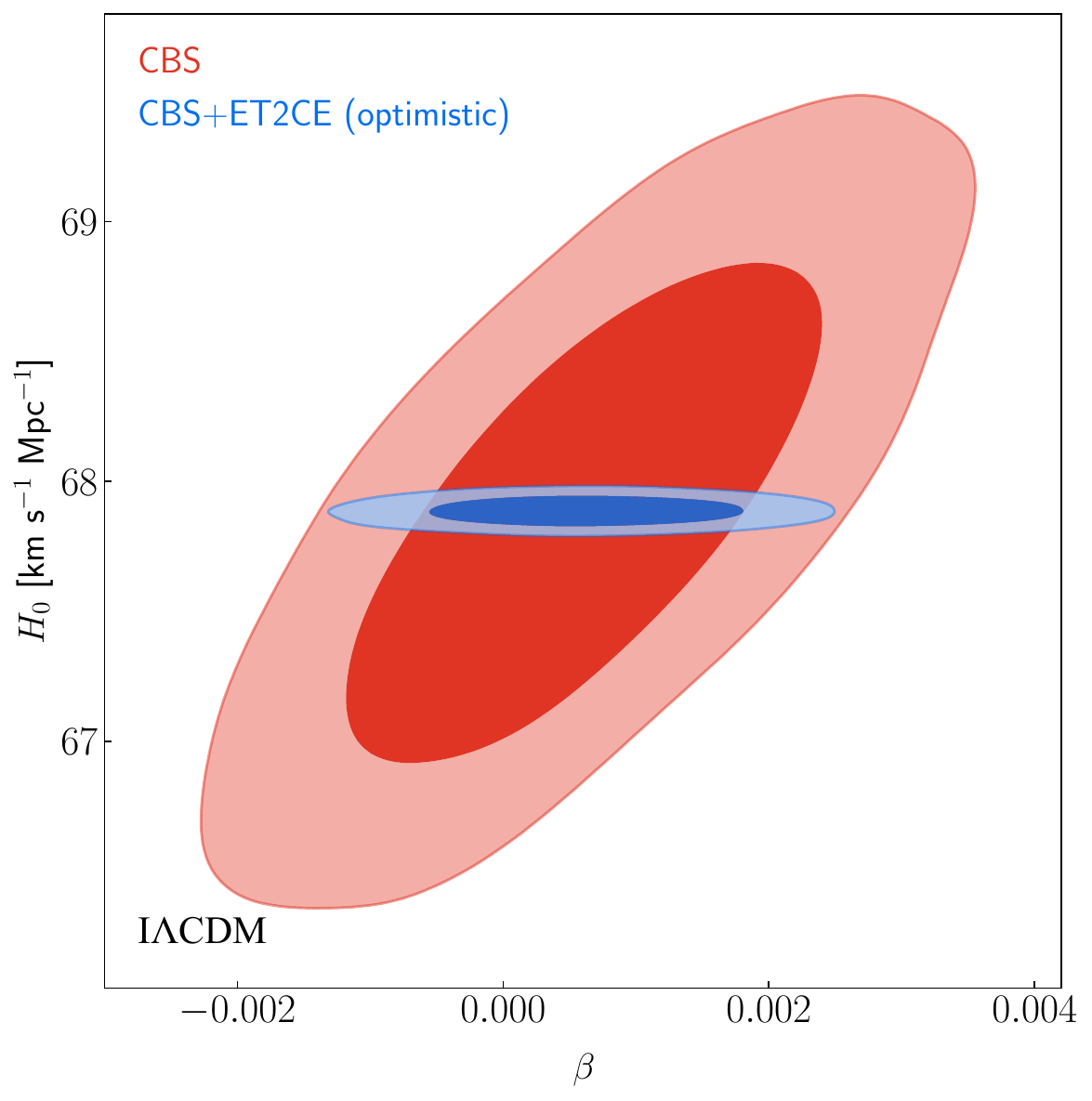}
	\includegraphics[width=0.45\linewidth,angle=0]{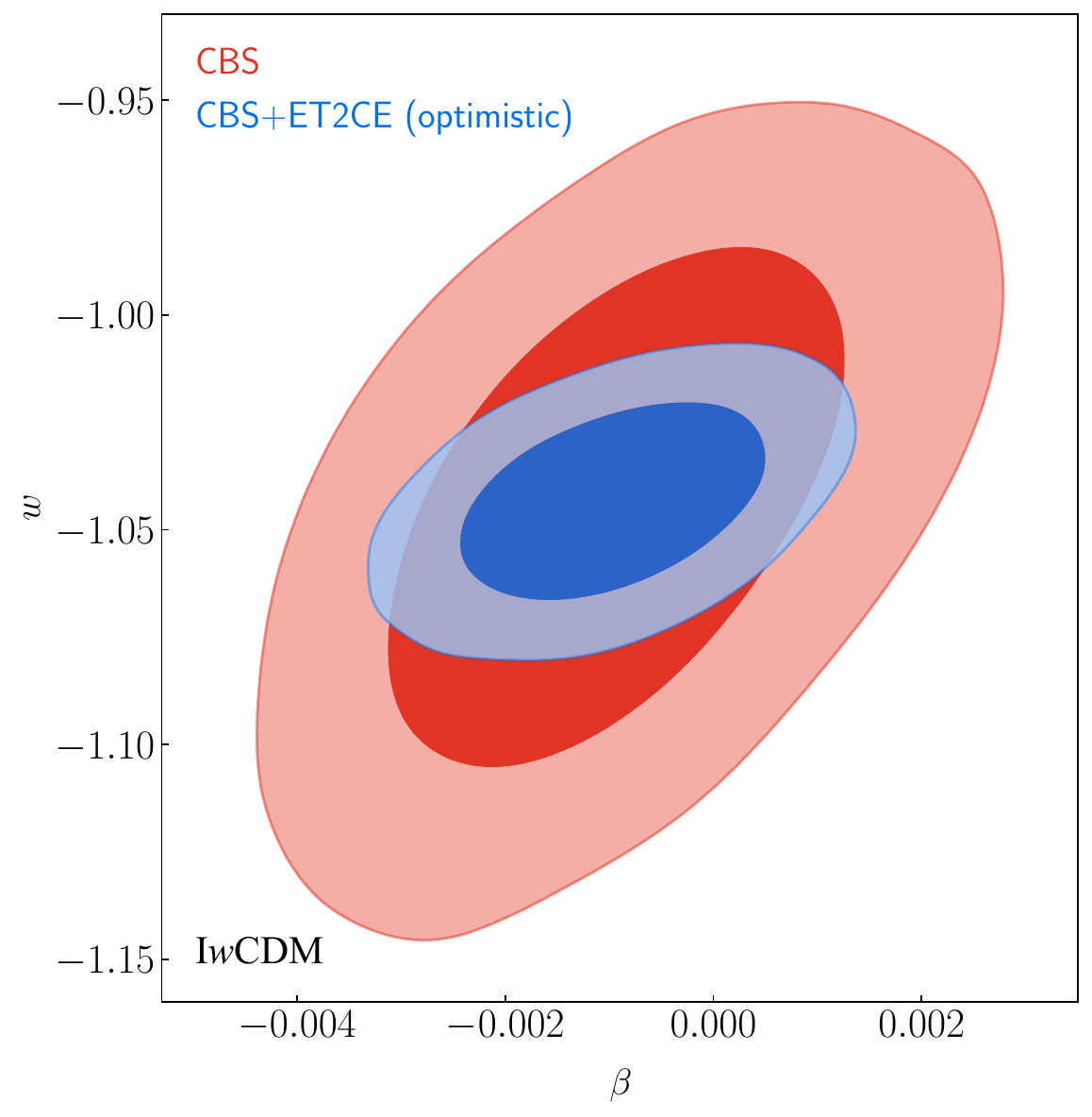}
	\caption{\label{fig12} Two-dimensional marginalized contours (68.3\% and 95.4\% confidence level) in the $\beta$--$H_0$ and $\beta$--$w$ planes using the CBS and CBS+ET2CE data in the optimistic scenario.}
\end{figure*}

In Fig.~\ref{fig11}, we show the constraint results for the $w$CDM and $w_0w_a$CDM models. The above constraint results of GW still hold, i.e., ET2CE gives the best constraint results, followed by 2CE, CE, and ET. For the constraint on $w$, ET2CE gives $\sigma(w)=0.045$, which is slightly worse that of CBS. In the case of the $w_0w_a$CDM model, ET2CE gives better constraint on $w_0$ and worse constraint on $w_a$ compared to the constraint results of CBS. In Fig.~\ref{fig12}, we also show the constraint results for the I$\Lambda$CDM and I$w$CDM models. With the addition of ET2CE to CBS, the constraints on cosmological parameters could be improved by 60.6\%--94.9\%, 59.4\%--92.4\%, 34.2\%--93.6\%, and 36.0\%--94.1\% in the $w$CDM, $w_0w_a$CDM, I$\Lambda$CDM, and I$w$CDM models, respectively. Moreover, CBS+ET2CE gives $\sigma(w)=0.013$ in the $w$CDM model with a precision of 1.26\%, which is close to the standard of precision cosmology. It is worth expecting that the fundamental nature of dark energy can be probed with the help of the 3G GW standard sirens.

\begin{table*}[!htb]
	\caption{The absolute (1$\sigma$) and relative errors of the cosmological parameters in the $\Lambda$CDM, $w$CDM, and $w_0w_a$CDM models using the ET, CE, 2CE, ET2CE, CBS, CBS+ET, CBS+CE, CBS+2CE, and CBS+ET2CE data in the optimistic scenario for the FOV. Here $H_0$ is in units of $\rm km\ s^{-1}\ Mpc^{-1}$. Note that $\sigma(\xi)$ and $\varepsilon(\xi)=\sigma(\xi)/\xi$ represent the absolute and relative errors of the parameter $\xi$, respectively.}
	\label{tab4}
	\setlength{\tabcolsep}{1.7mm}
	\renewcommand{\arraystretch}{1.5}
	\begin{center}{\centerline{
				\begin{tabular}{{|c|c|c|c|c|c|c|c|c|c|c|}}
					\hline   
					Model       & Error                           &ET               &CE                &2CE               &ET2CE          &CBS            &CBS+ET        &CBS+CE      &CBS+2CE      &CBS+ET2CE \\ \hline
					\multirow{4}{*}{$\Lambda$CDM}  
					&$\sigma(\Omega_{\rm m})$       &$0.0130$      &$0.0072$    &$0.0062$    &$0.0053$     &$0.0059$   &$0.0018$    &$0.0011$     &$0.0011$     &$0.0011$      \\
					&$\sigma(H_0)$                            &$0.140$        &$0.052$      &$0.047$       &$0.044$        &$0.440$       &$0.110$     &$0.049$      &$0.044$      &$0.041$        \\
					&$\ve(\Omega_{\rm m})$             &$4.18\%$       &$2.32\%$  &$2.00\%$   &$1.71\%$       &$1.90\%$  &$0.58\%$       &$0.35\%$  &$0.35\%$      &$0.35\%$    \\
					&$\ve(H_0)$                                  &$0.207\%$     &$0.077\%$  &$0.069\%$  &$0.065\%$   &$0.651\%$ &$0.163\%$    &$0.072\%$   &$0.065\%$   &$0.061\%$   \\  \hline
					\multirow{6}{*}{$w$CDM}
					&$\sigma(\Omega_{\rm m})$       &$0.0435$   &$0.0240$      &$0.0200$      &$0.0160$    &$0.0076$  &$0.0022$    &$0.0020$    &$0.0019$     &$0.0018$      \\
					&$\sigma(H_0)$                            &$0.170$        &$0.055$        &$0.051$        &$0.048$        &$0.820$       &$0.120$        &$0.050$       &$0.045$       &$0.042$       \\
					&$\sigma(w)$                                &$0.120$      &$0.064$      &$0.053$       &$0.045$           &$0.033$     &$0.018$      &$0.015$       &$0.014$        &$0.013$        \\
					&$\ve(\Omega_{\rm m})$             &$14.22\%$   &$7.82\%$   &$6.51\%$   &$5.21\%$         &$2.48\%$ &$0.72\%$  &$0.65\%$ &$0.62\%$   &$0.59\%$   \\
					&$\ve(H_0)$                                  &$0.249\%$ &$0.081\%$  &$0.075\%$    &$0.070\%$    &$1.201\%$  &$0.176\%$  &$0.073\%$  &$0.066\%$   &$0.062\%$   \\
					&$\ve(w)$                                      &$11.54\%$ &$6.17\%$     &$5.12\%$   &$4.36\%$                &$3.20\%$ &$1.75\%$  &$1.46\%$  &$1.36\%$   &$1.26\%$    \\ \hline
					\multirow{7}{*}{$w_0w_a$CDM}
					&$\sigma(\Omega_{\rm m})$       &$0.0830$   &$0.0660$    &$0.0625$     &$0.0565$       &$0.0077$    &$0.0030$    &$0.0027$   &$0.0026$      &$0.0026$     \\
					&$\sigma(H_0)$                            &$0.235$     &$0.102$       &$0.091$          &$0.087$        &$0.820$     &$0.150$    &$0.073$        &$0.065$           &$0.062$       \\
					&$\sigma(w_0)$                            &$0.120$      &$0.075$      &$0.067$        &$0.057$      &$0.082$     &$0.044$      &$0.030$     &$0.027$        &$0.024$        \\
					&$\sigma(w_a)$                            &$1.59$      &$0.95$       &$0.88$              &$0.78$          &$0.32$        &$0.19$        &$0.15$        &$0.14$      &$0.13$          \\
					&$\ve(\Omega_{\rm m})$             &$25.94\%$&$22.92\%$&$22.08\%$&$20.47\%$       &$2.50\%$       &$0.98\%$ &$0.88\%$ &$0.85\%$    &$0.85\%$   \\
					&$\ve(H_0)$                                  &$0.345\%$ &$0.149\%$   &$0.133\%$   &$0.127\%$   &$1.201\%$   &$0.220\%$ &$0.107\%$  &$0.095\%$    &$0.091\%$     \\
					&$\ve(w_0)$                                  &$12.90\%$&$8.13\%$  &$7.26\%$   &$6.20\%$             &$8.61\%$   &$4.62\%$   &$3.15\%$    &$2.84\%$   &$2.52\%$    \\  \hline			
		\end{tabular}}}
	\end{center}
\end{table*}

\begin{table*}[!htb]
	\caption{The absolute (1$\sigma$) and relative errors of the cosmological parameters in the I$\Lambda$CDM and I$w$CDM models using the CBS, CBS+ET, CBS+CE, CBS+2CE, and CBS+ET2CE data in the optimistic scenario for the FOV.}
	\label{tab5}
	\setlength{\tabcolsep}{2.15mm}
	\renewcommand{\arraystretch}{1.5}
	\begin{center}{\centerline{
				\begin{tabular}{|c|c|m{2.5cm}<{\centering}|m{2.5cm}<{\centering}|m{2.5cm}<{\centering}|m{2.5cm}<{\centering}|m{2.5cm}<{\centering}|}
					\hline   
					Model       & Error                                 &CBS            &CBS+ET        &CBS+CE      &CBS+2CE      &CBS+ET2CE \\ \hline
					\multirow{5}{*}{I$\Lambda$CDM}
					&$\sigma(\Omega_{\rm m})$          &$0.0081$   &$0.0018$   &$0.0012$    &$0.0012$     &$0.0011$       \\
					&$\sigma(H_0)$                               &$0.640$       &$0.110$         &$0.049$      &$0.045$       &$0.041$         \\
					&$\sigma(\beta)$                           &$0.00120$    &$0.00080$&$0.00079$  &$0.00079$  &$0.00079$      \\
					&$\ve(\Omega_{\rm m})$             &$2.63\%$    &$0.58\%$  &$0.39\%$&$0.39\%$  &$0.36\%$    \\
					&$\ve(H_0)$                                   &$0.943\%$    &$0.162\%$   &$0.072\%$ &$0.066\%$  &$0.060\%$      \\ \hline
					\multirow{7}{*}{I$w$CDM}
					&$\sigma(\Omega_{\rm m})$       &$0.0080$   &$0.0023$   &$0.0020$   &$0.0019$     &$0.0018$       \\
					&$\sigma(H_0)$                           &$0.820$         &$0.140$       &$0.056$      &$0.051$       &$0.048$        \\
					&$\sigma(w)$                               &$0.040$      &$0.025$    &$0.019$       &$0.017$       &$0.015$         \\
					&$\sigma(\beta)$                         &$0.00150$     &$0.00120$  &$0.00100$     &$0.00100$     &$0.00096$    \\
					&$\ve(\Omega_{\rm m})$            &$2.60\%$     &$0.75\%$   &$0.65\%$  &$0.62\%$   &$0.58\%$    \\
					&$\ve(H_0)$                                  &$1.202\%$   &$0.205\%$   &$0.082\%$      &$0.075\%$        &$0.070\%$    \\
					&$\ve(w)$                                   &$3.83\%$   &$2.39\%$&$1.82\%$  &$1.63\%$   &$1.44\%$     \\  \hline				
		\end{tabular}}}
	\end{center}
\end{table*}

\subsection{Constraint results in realistic scenario}

\begin{figure*}[htbp]
	\includegraphics[width=0.45\linewidth,angle=0]{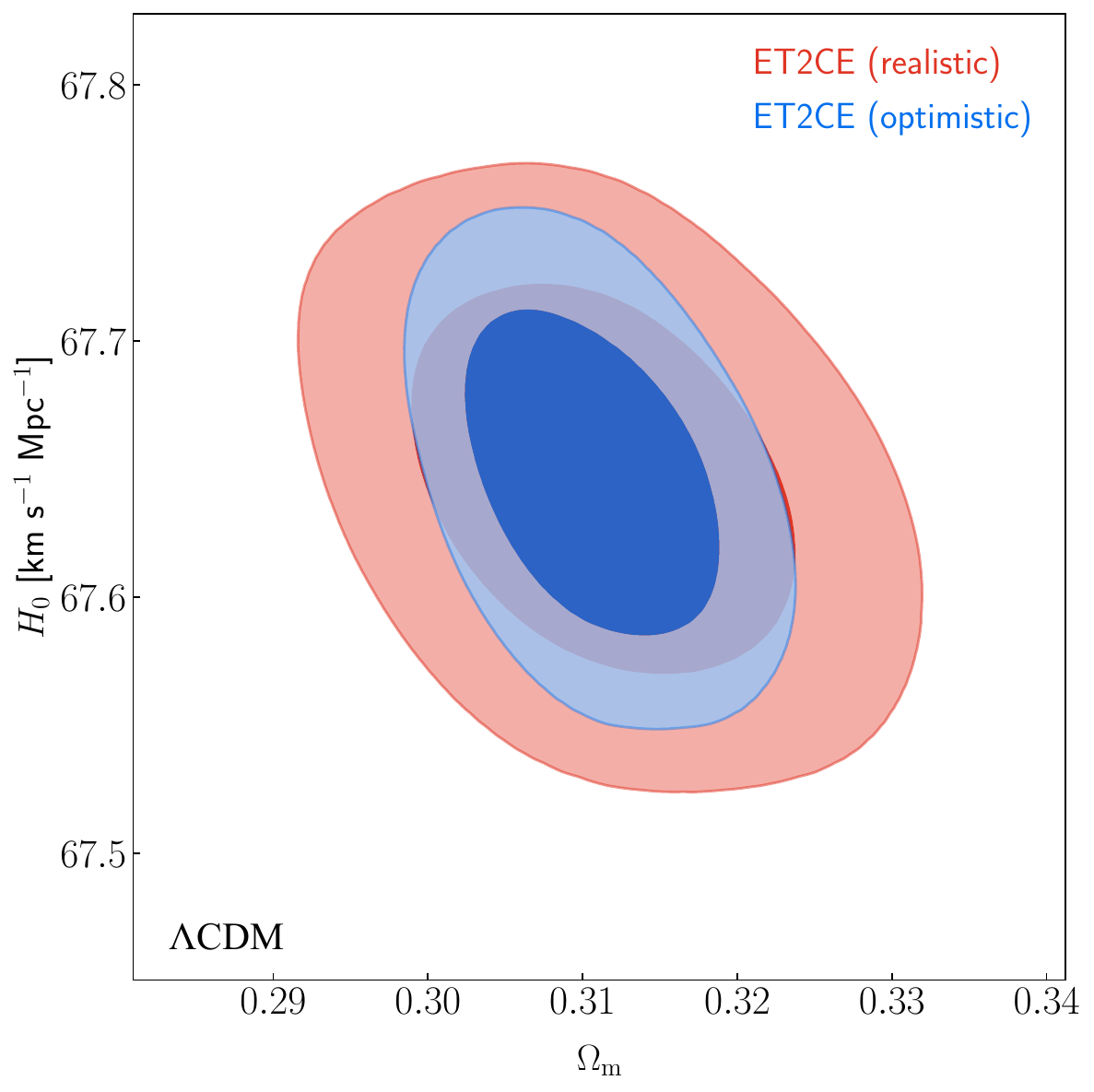}
    \includegraphics[width=0.45\linewidth,angle=0]{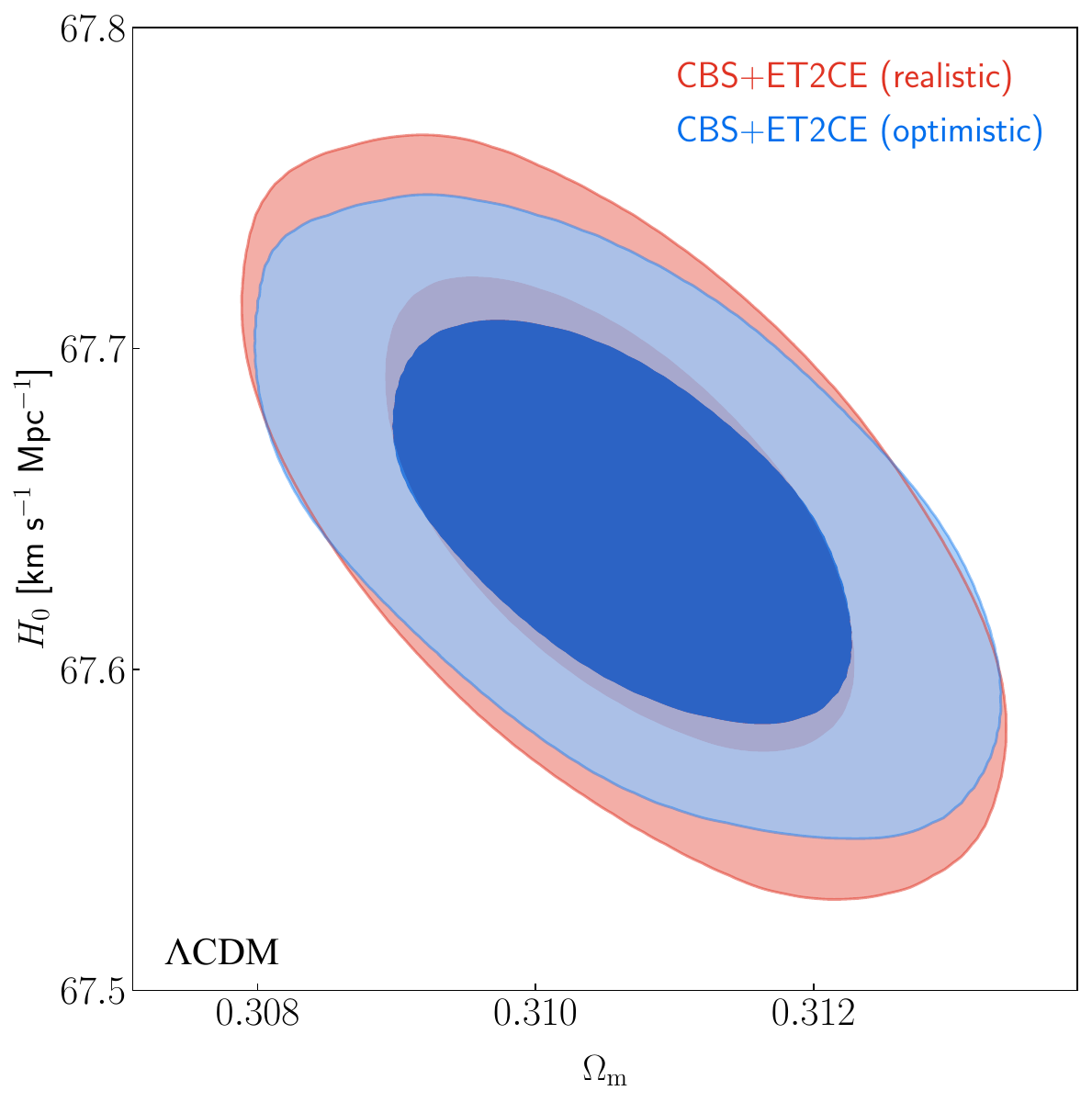}
	\caption{\label{fig13} Constraints on the $\Lambda$CDM model in the realistic and optimistic scenarios. Left panel: Two-dimensional marginalized contours (68.3\% and 95.4\% confidence level) in the $\Omega_{\rm m}$--$H_0$ plane using the realistic and optimistic scenarios of the ET2CE data. Right panel: Two-dimensional marginalized contours (68.3\% and 95.4\% confidence level) in the $\Omega_{\rm m}$--$H_0$ plane using the CBS+ET2CE (realistic) and CBS+ET2CE (optimistic) data.}
\end{figure*}

\begin{figure*}[htbp]
    \includegraphics[width=0.45\linewidth,angle=0]{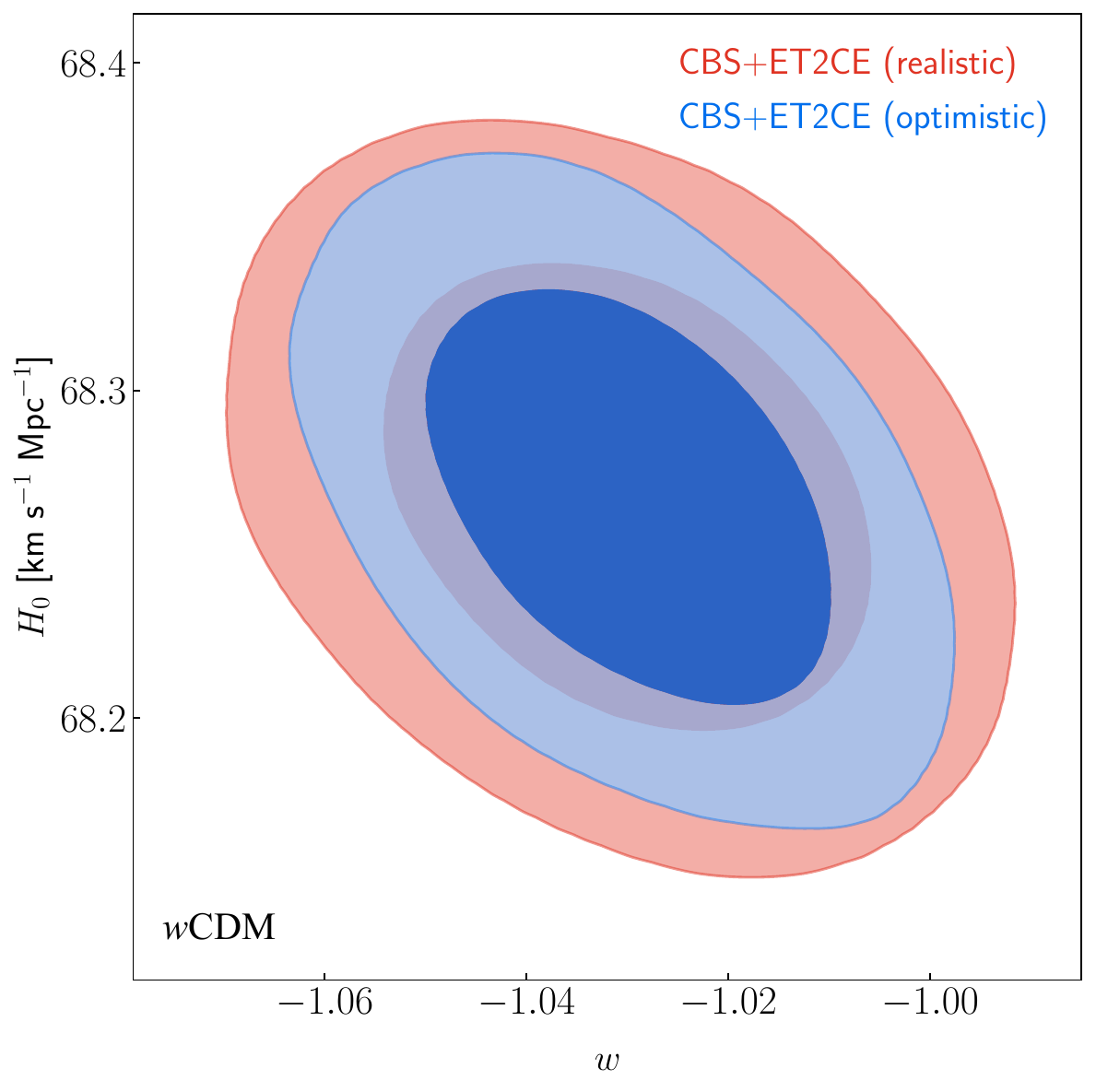}
	\includegraphics[width=0.45\linewidth,angle=0]{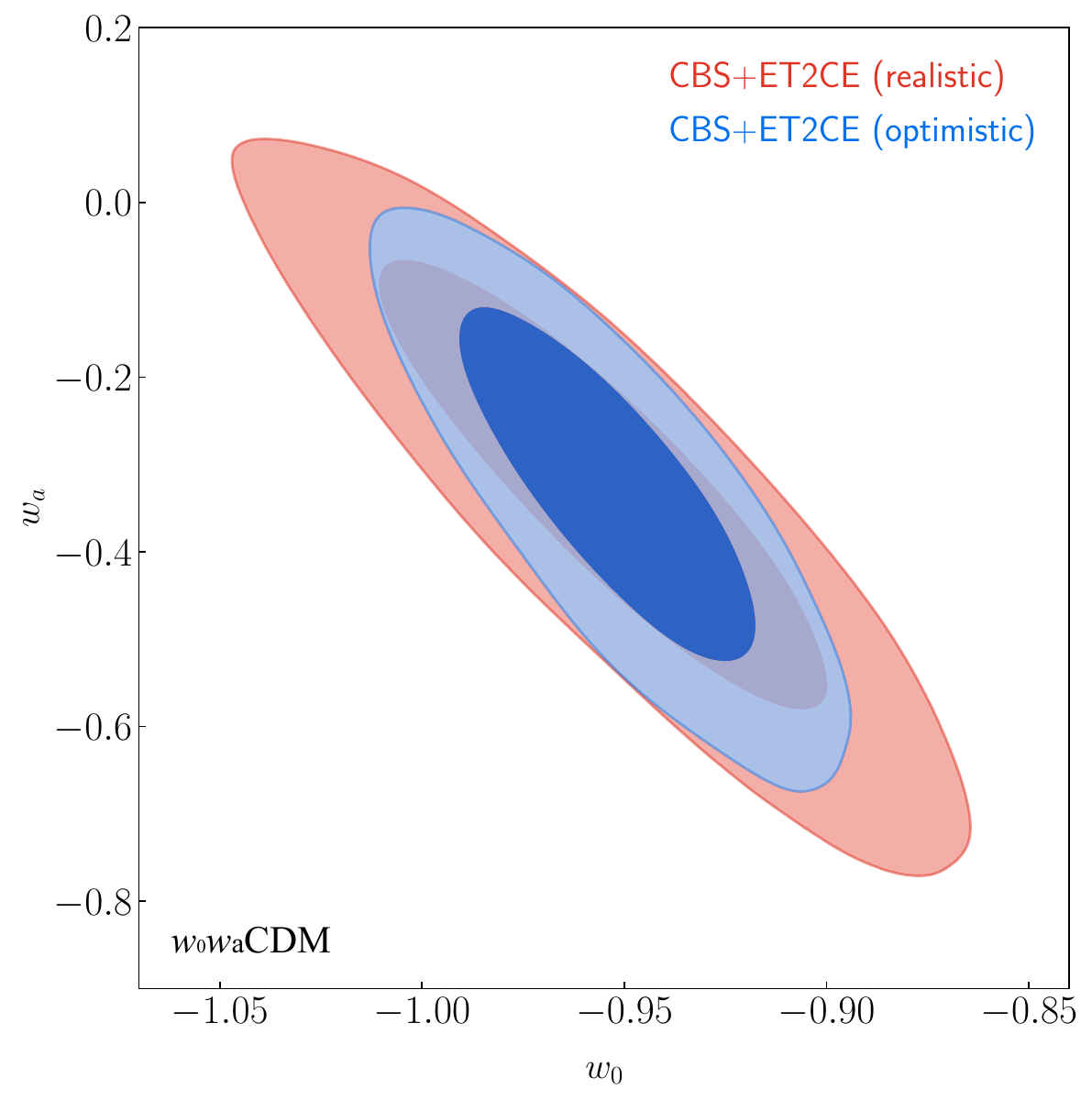}
    \includegraphics[width=0.45\linewidth,angle=0]{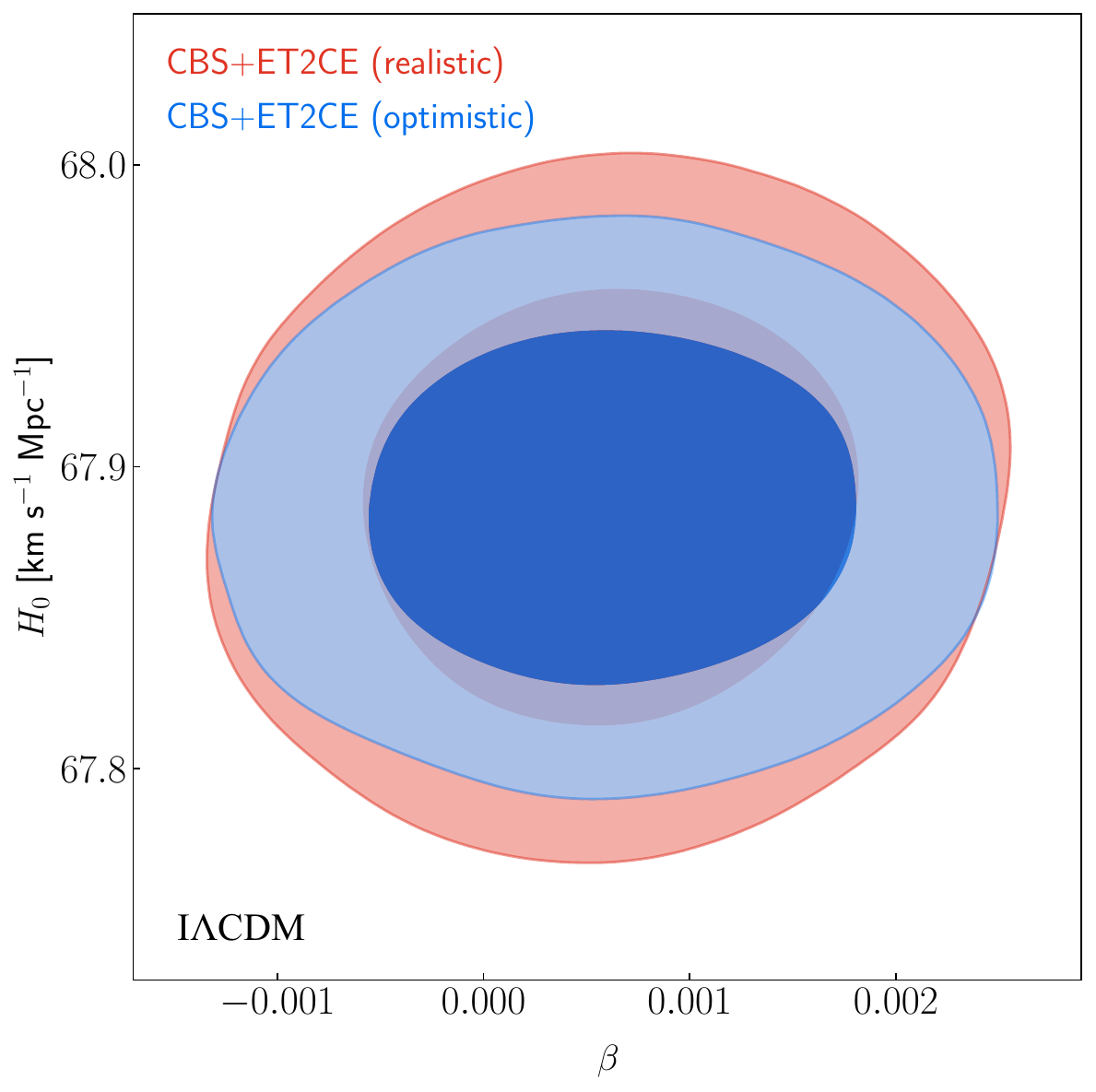}
    \includegraphics[width=0.45\linewidth,angle=0]{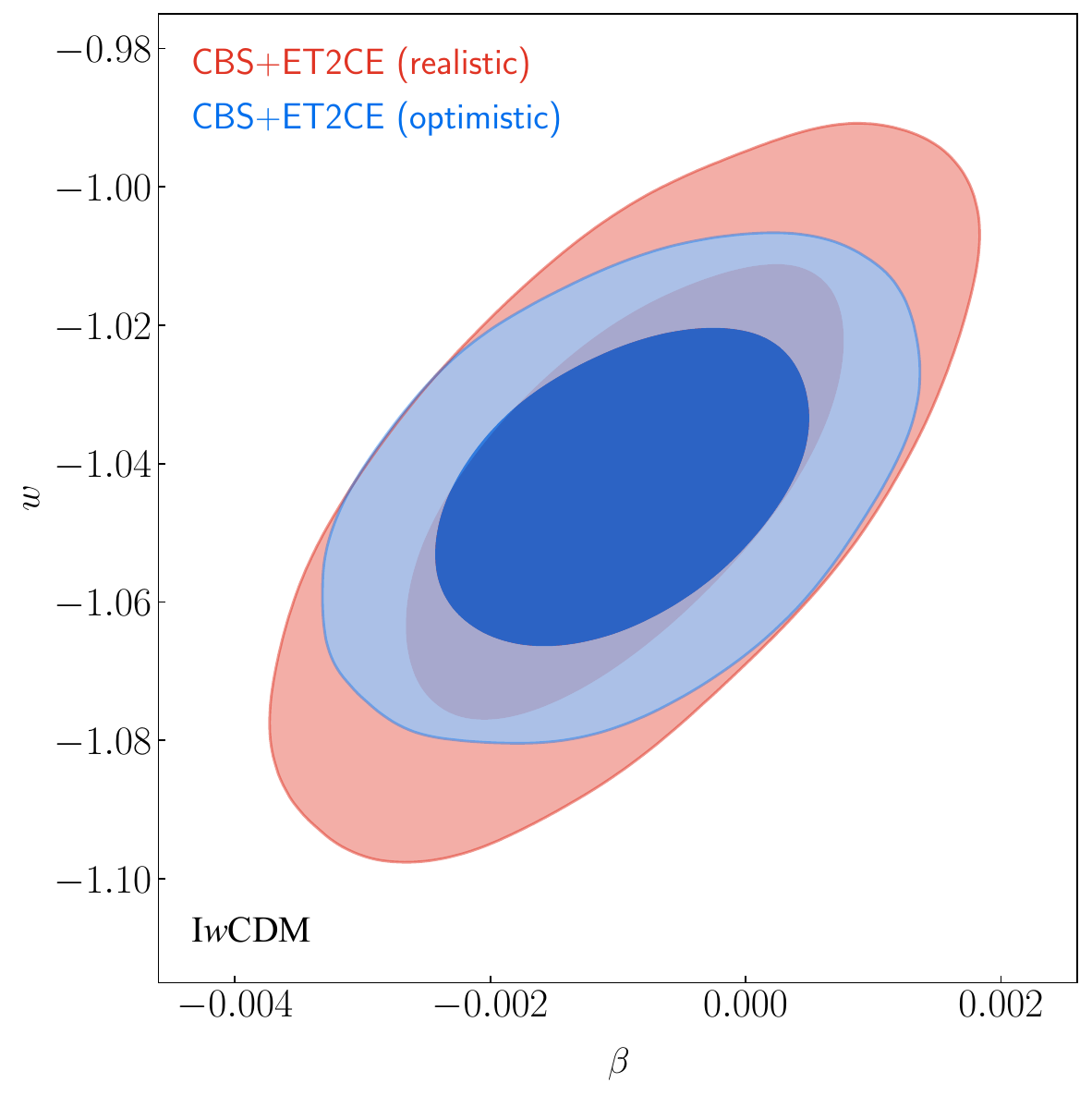}
	\caption{\label{fig14} Two-dimensional marginalized contours (68.3\% and 95.4\% confidence level) in the $w$--$H_0$, $w_0$--$w_a$, $\beta$--$H_0$, and $\beta$--$w$ planes using the CBS+ET2CE (realistic) and CBS+ET2CE (optismistic) data.}
\end{figure*}

In Fig.~\ref{fig13}, we show the constraint results in the $\Omega_{\rm m}$--$H_0$ planes for the $\Lambda$CDM model. In the left panel of Fig.~\ref{fig13}, we could see that ET2CE (optimistic) gives better constraint results than those of ET2CE (realistic). Concretely, ET2CE (realistic) gives $\sigma(\Omega_{\rm m})=0.0088$ and $\sigma(H_0)=0.051$ which is 66.0\% and 15.9\% worse than those of ET2CE (optimistic). In the right panel of Fig.~\ref{fig13}, we can see that CBS+ET2CE (optimistic) also gives better constraints on cosmological parameters than those of CBS+ET2CE (realistic), although the difference is mitigated. CBS+ET2CE (realistic) gives $\sigma(H_0)=0.049$ which is 19.5\% worse than those of CBS+ET2CE (optimistic). In Fig.~\ref{fig14}, we show the cases for the $w$CDM, $w_0w_a$CDM, I$\Lambda$CDM, and I$w$CDM models. We can also see that CBS+ET2CE (optimistic) gives better constraints on cosmological parameters than those of CBS+ET2CE (realistic). 

\begin{table*}[!htb]
	\caption{Same as in Table~\ref{tab4}, assuming the realistic scenario for the FOV.}
	\label{tab6}
	\setlength{\tabcolsep}{1.7mm}
	\renewcommand{\arraystretch}{1.5}
	\begin{center}{\centerline{
				\begin{tabular}{{|c|c|c|c|c|c|c|c|c|c|c|}}
					\hline   
					Model       & Error                           &ET               &CE                &2CE               &ET2CE          &CBS            &CBS+ET        &CBS+CE      &CBS+2CE      &CBS+ET2CE \\ \hline
					\multirow{4}{*}{$\Lambda$CDM}  
					&$\sigma(\Omega_{\rm m})$       &$0.0230$      &$0.0120$      &$0.0100$       &$0.0088$     &$0.0059$   &$0.0023$    &$0.0012$     &$0.0011$     &$0.0011$      \\
					&$\sigma(H_0)$                            &$0.190$        &$0.059$        &$0.053$          &$0.051$        &$0.440$       &$0.160$  &$0.057$         &$0.051$      &$0.049$         \\
					&$\ve(\Omega_{\rm m})$             &$7.37\%$  &$3.86\%$      &$3.22\%$         &$2.83\%$   &$1.90\%$ &$0.74\%$  &$0.39\%$  &$0.35\%$  &$0.35\%$      \\
					&$\ve(H_0)$                                  &$0.281\%$  &$0.087\%$ &$0.078\%$   &$0.075\%$   &$0.651\%$ &$0.237\%$  &$0.084\%$   &$0.075\%$  &$0.072\%$       \\  \hline
					\multirow{6}{*}{$w$CDM}
					&$\sigma(\Omega_{\rm m})$       &$0.0835$    &$0.0390$      &$0.0340$      &$0.0275$       &$0.0076$  &$0.0026$    &$0.0022$    &$0.0022$      &$0.0021$      \\
					&$\sigma(H_0)$                            &$0.200$        &$0.060$          &$0.055$      &$0.052$          &$0.820$   &$0.160$        &$0.057$    &$0.052$          &$0.049$           \\
					&$\sigma(w)$                                &$0.215 $     &$0.103$       &$0.092$          &$0.076$     &$0.033$     &$0.019$      &$0.017$       &$0.017$        &$0.016$        \\
					&$\ve(\Omega_{\rm m})$             &$27.20\%$&$12.70\%$  &$11.07\%$&$8.93\%$   &$2.48\%$ &$0.85\%$  &$0.72\%$ &$0.72\%$   &$0.69\%$    \\
					&$\ve(H_0)$                                  &$0.293\%$  &$0.088\%$  &$0.081\%$  &$0.076\%$   &$1.201\%$ &$0.234\%$  &$0.083\%$ &$0.076\%$   &$0.072\%$     \\
					&$\ve(w)$                                      &$20.09\%$&$9.90\%$&$8.85\%$   &$7.31\%$   &$3.20\%$&$1.84\%$  &$1.65\%$  &$1.65\%$   &$1.55\%$    \\ \hline
					\multirow{7}{*}{$w_0w_a$CDM}
					&$\sigma(\Omega_{\rm m})$       &$0.0940$       &$0.0860$   &$0.0810$    &$0.0730$      &$0.0077$    &$0.0031$   &$0.0027$   &$0.0026$     &$0.0026$     \\
					&$\sigma(H_0)$                            &$0.260$         &$0.125$        &$0.117$      &$0.104$          &$0.820$        &$0.180$       &$0.080$       &$0.072$           &$0.068$          \\
					&$\sigma(w_0)$                            &$0.195$       &$0.110$        &$0.100$          &$0.085$           &$0.082$     &$0.059$     &$0.045$     &$0.040$        &$0.036$       \\
					&$\sigma(w_a)$                            &$2.40$         &$1.43$         &$1.28$        &$1.05$            &$0.32$       &$0.25$       &$0.20$        &$0.18$          &$0.17$        \\
					&$\ve(\Omega_{\rm m})$             &$26.86\%$  &$28.86\%$   &$27.00\%$&$25.17\%$  &$2.50\%$ &$1.01\%$ &$0.88\%$  &$0.85\%$    &$0.85\%$    \\
					&$\ve(H_0)$                                  &$0.381\%$   &$0.183\%$  &$0.172\%$  &$0.152\%$    &$1.201\%$  &$0.264\%$ &$0.117\%$ &$0.105\%$   &$0.100\%$   \\
					&$\ve(w_0)$                                  &$20.10\%$    &$12.09\%$     &$10.87\%$ &$9.25\%$  &$8.61\%$  &$6.18\%$ &$4.72\%$  &$4.19\%$   &$3.78\%$    \\  \hline
			\end{tabular}}}
	\end{center}
\end{table*}

\begin{table*}[!htb]
	\caption{Same as in Table~\ref{tab5}, assuming the realistic scenario for the FOV.}
	\label{tab7}
	\setlength{\tabcolsep}{2.15mm}
	\renewcommand{\arraystretch}{1.5}
	\begin{center}{\centerline{
				\begin{tabular}{|c|c|m{2.5cm}<{\centering}|m{2.5cm}<{\centering}|m{2.5cm}<{\centering}|m{2.5cm}<{\centering}|m{2.5cm}<{\centering}|}
					\hline   
					Model       & Error                                 &CBS            &CBS+ET        &CBS+CE      &CBS+2CE      &CBS+ET2CE \\ \hline
					\multirow{5}{*}{I$\Lambda$CDM}
					&$\sigma(\Omega_{\rm m})$       &$0.0081$   &$0.0023$    &$0.0012$    &$0.0012$     &$0.0012$       \\
					&$\sigma(H_0)$                            &$0.640$       &$0.160$         &$0.057$         &$0.052$         &$0.049$         \\
					&$\sigma(\beta)$                           &$0.00120$    &$0.00081$  &$0.00080$ &$0.00080$  &$0.00079$    \\
					&$\ve(\Omega_{\rm m})$              &$2.63\%$  &$0.75\%$  &$0.39\%$  &$0.39\%$  &$0.39\%$    \\
					&$\ve(H_0)$                                  &$0.943\%$ &$0.236\%$  &$0.084\%$ &$0.077\%$   &$0.072\%$     \\ \hline
					\multirow{7}{*}{I$w$CDM}
					&$\sigma(\Omega_{\rm m})$        &$0.0080$   &$0.0028$   &$0.0023$   &$0.0022$     &$0.0022$       \\
					&$\sigma(H_0)$                             &$0.820$        &$0.180$       &$0.061$        &$0.056$         &$0.054$            \\
					&$\sigma(w)$                                 &$0.040$      &$0.029$    &$0.024$       &$0.023$      &$0.022$         \\
					&$\sigma(\beta)$                          &$0.00150$     &$0.00130$  &$0.00120$     &$0.00120$    &$0.00110$         \\
					&$\ve(\Omega_{\rm m})$            &$2.60\%$  &$0.91\%$&$0.75\%$  &$0.71\%$   &$0.71\%$    \\
					&$\ve(H_0)$                                   &$1.202\%$   &$0.264\%$&$0.089\%$ &$0.082\%$  &$0.079\%$       \\
					&$\ve(w)$                                     &$3.83\%$   &$2.78\%$&$2.30\%$  &$2.20\%$  &$2.11\%$     \\ \hline			
		\end{tabular}}}
	\end{center}
\end{table*}

In the next decades, other powerful cosmological probes, 21 cm intensity mapping, fast radio bursts, and strong gravitational lensing, can also play a crucial role in exploring the evolution of the universe \cite{Wu:2022jkf,2023SCPMA..6670431C,Zhang:2023gye,2023SCPMA..6620431D,Walters:2017afr,SKA:2018ckk,Zhang:2021yof,Wu:2021vfz,Zhao:2022bpd,Qiu:2021cww,Zhao:2020ole,Futamase:2000hnr,Grillo:2007iv,Treu:2016ljm,Wong:2019kwg,Wang:2019yob,Qi:2020rmm,Qi:2022sxm,Qi:2022kfg,Liu:2021xvc,Wang:2021kxc,Wang:2022rvf}. Moreover, the synergies between GW and other cosmological probes are also discussed in Refs.~\cite{Jin:2021pcv,Wu:2022dgy}. A comprehensive discussion of these aspects will be made in our future works.

\subsection{Comparison with previous works}

Compared to previous works~\cite{Zhang:2019ylr,Zhang:2019loq,Zhang:2018byx,Li:2019ajo,Jin:2020hmc,Jin:2021pcv,Zhang:2019ple,Jin:2023zhi,Wu:2022dgy}, our main differences in this paper are as follows.

Previous works roughly assume 1000 standard sirens with detectable EM counterparts for ET or CE alone in a 10-year observation. Such treatment is optimistic and may not be realistic. In fact, in the case of a single ET with the optimistic estimate, we have only about 370 detectable GW-GRB events. Even in the case of the ET2CE network, the number of detectable GW-GRB events is about 620 in the present work. It is seen that the number of standard sirens has been significantly overestimated in the previous works. In Ref.~\cite{Hou:2022rvk}, they estimated about 400 standard sirens for a single ET in a 10-year observation, which gave a slightly higher number of standard sirens. The prime cause is that we adopt the SNR threshold to be 12, while Hou~\emph{et al.}~\cite{Hou:2022rvk} adopt the SNR threshold to be 8.

Previous works roughly assumed that the estimated standard sirens directly followed the distribution in redshift determined by the star formation rate with long tails at larger redshift. However, in the joint GW-GRB observations, we cannot ignore the influence of the detection rate of GW detectors and GRB detectors for the events with redshifts. This leads to a lower redshift distribution, mainly at $z\in [0,2]$ in Figs.~\ref{fig5} and~\ref{fig6}, instead of the range of $z\in[0,5]$.

Compared to previous works, it could be surprising that for a single GW detector, our constraint results of $H_0$ are tighter. The main reason is that the redshift distributions of the standard sirens are lower as mentioned above.

In Ref.~\cite{Hou:2022rvk}, they only focused on the synergy of ET alone with the THESEUS mission in the optimistic scenario for the GW-GRB detection in the IDE models. However, we make a comprehensive analysis of four different cases of 3G GW observations, single ET, single CE, the 2CE network, and the ET2CE network. Moreover, in this paper, we use the Fisher information matrix to estimate the instrumental error of the luminosity distance instead of using the approximation $2d_{\rm L}/\rho$. In addition, we also consider the realistic scenario for the GW-GRB detection. For the IDE models, we employ the ePPF method to avoid the cosmological perturbations. Compared to Ref.~\cite{Hou:2022rvk}, with these improvements above, it is strongly convinced that our results could better show the potential of the cosmological parameter estimations using the 3G-era standard sirens.

\subsection{Impact of the mass distributions of NSs on cosmological analysis}

In this subsection, we investigate the impact of the mass distributions of NSs on cosmological analysis. Here we choose four typical mass distributions of NSs, i.e., the Galactic BNS mass distribution~\cite{Ozel:2016oaf}, the Galactic NS mass distribution~\cite{Farr_2020}, the POWER population model~\cite{KAGRA:2021duu}, and the PEAK population model~\cite{KAGRA:2021duu}, to perform cosmological analysis. For the mass distribution of the latter three types of NSs, we employ the numerical fitting formulas to fit the curves shown in Fig.~7 of Ref.~\cite{KAGRA:2021duu}. Note that the following discussions are based on the $\Lambda$CDM model using ET and ET2CE in the optimistic scenario.

\begin{figure*}[htbp]
	\includegraphics[width=0.7\linewidth,angle=0]{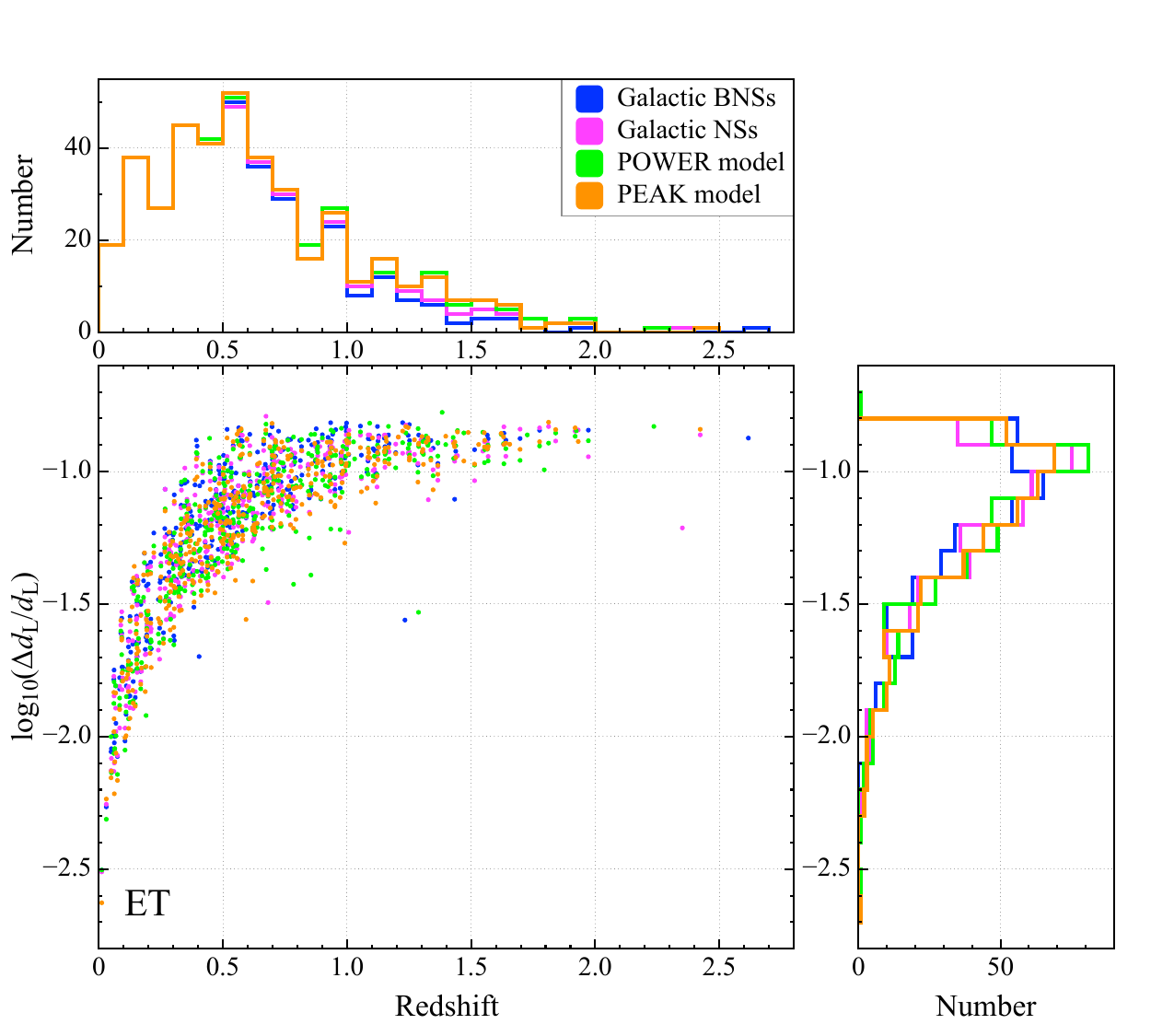}
	\includegraphics[width=0.7\linewidth,angle=0]{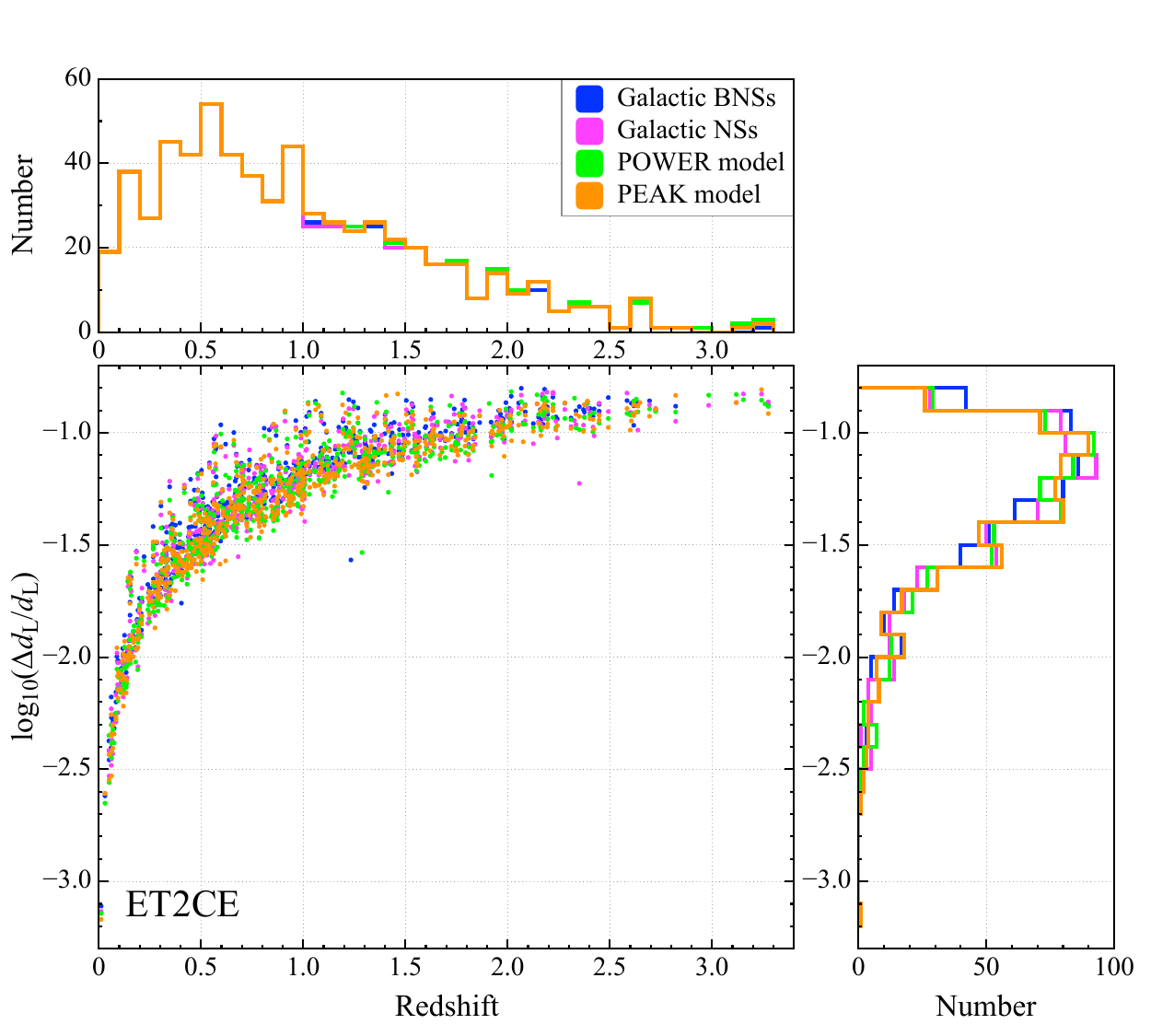}
	\caption{\label{fig15} Distributions of $\Delta d_{\rm L}/d_{\rm L}$ with respective to the redshifts for GW standard sirens using four mass distributions of NSs in the optimistic scenario. The upper panel shows the results of ET and the lower panel shows those of ET2CE. In each panel, the blue, purple, green, and orange dots and histograms indicate the Galactic BNS mass distribution, the Galactic NS mass distribution, the POWER population model, and the PEAK population model, respectively.}
\end{figure*}

\begin{figure}[htbp]
	\includegraphics[width=0.9\linewidth,angle=0]{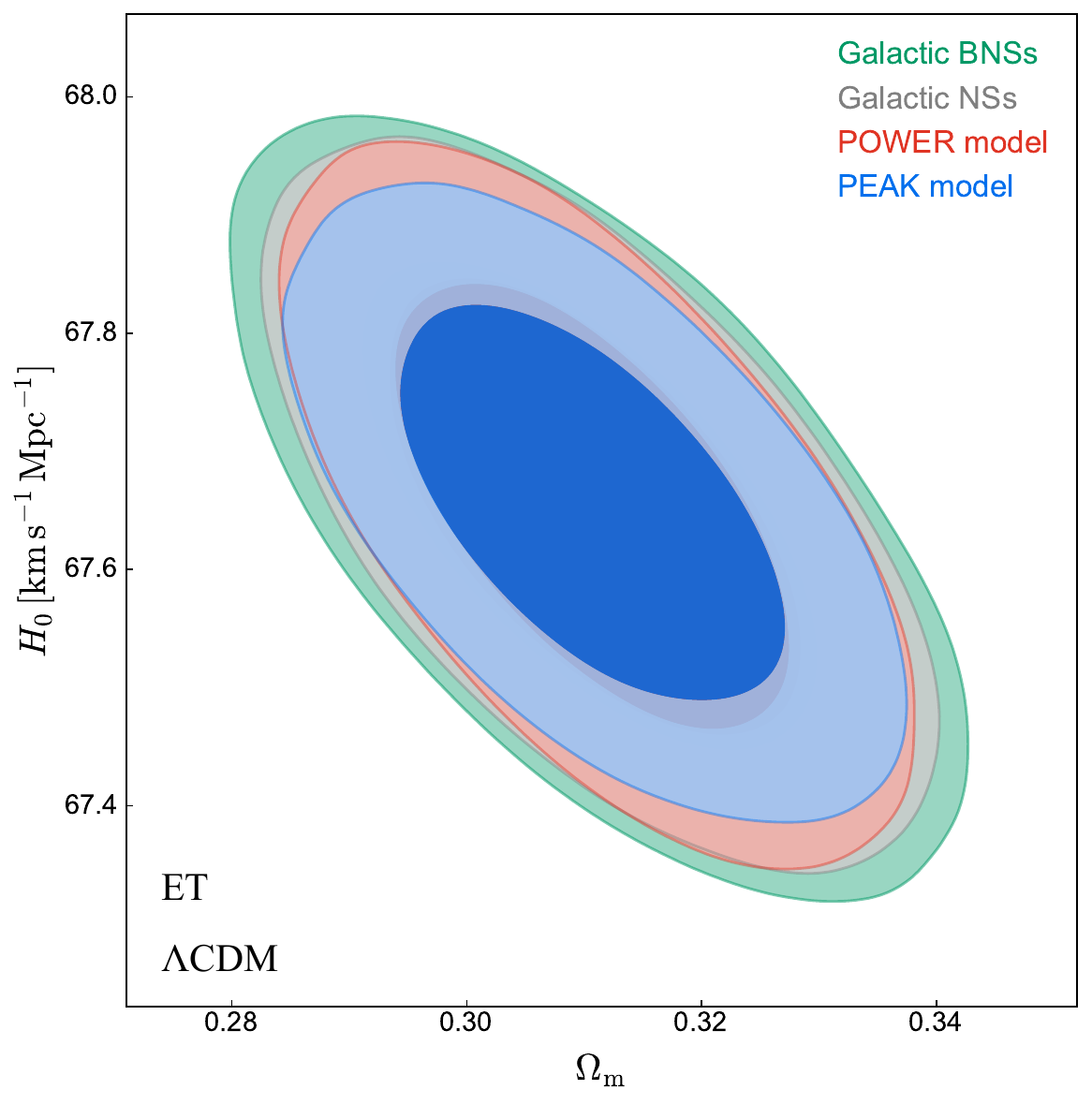}
	\includegraphics[width=0.9\linewidth,angle=0]{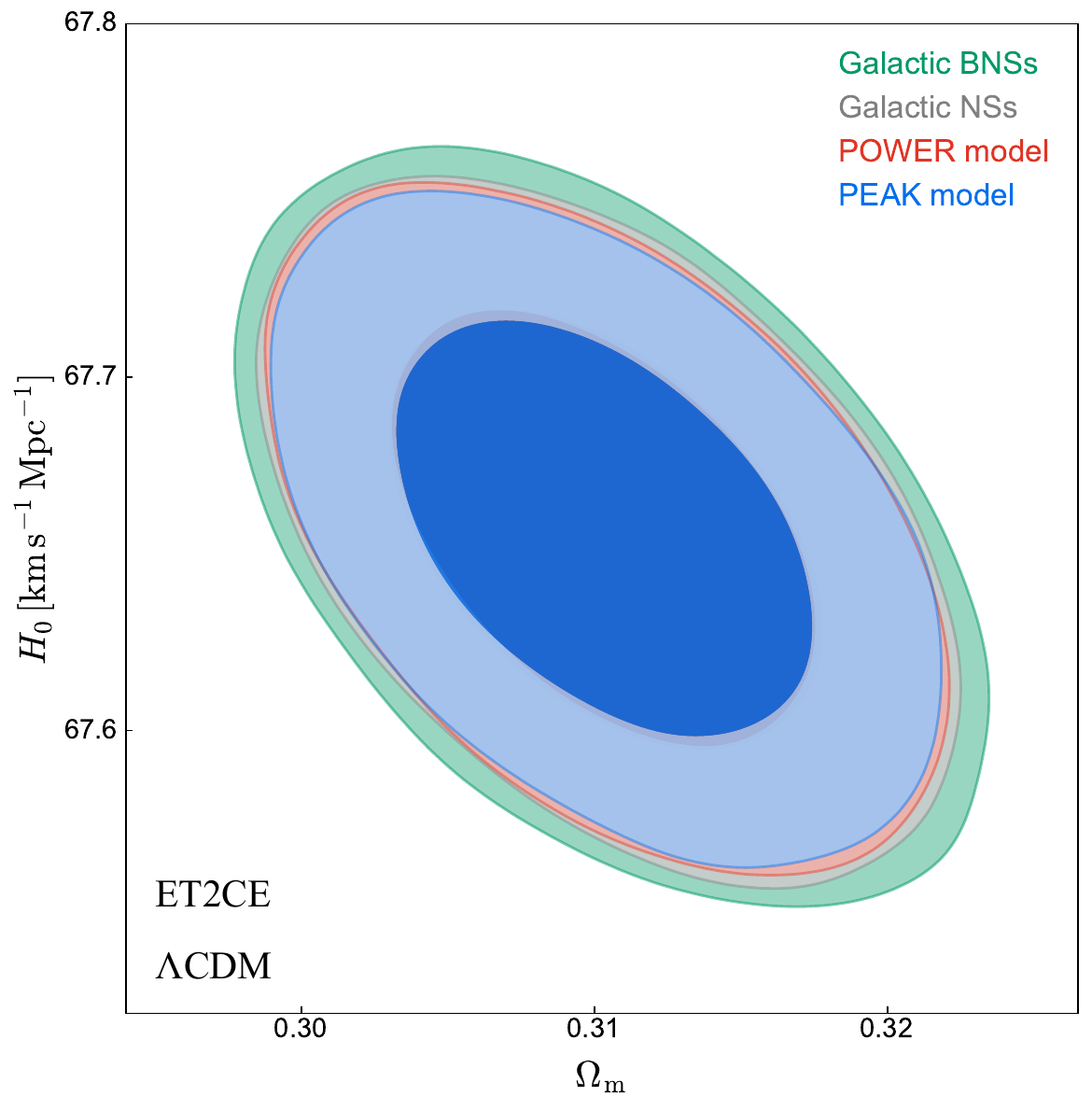}
	\caption{\label{fig16} Constraints on the $\Lambda$CDM model using four mass distributions of NSs in the optimistic scenario. Upper panel: Two-dimensional marginalized contours (68.3\% and 95.4\% confidence level) in the $\Omega_{\rm m}$--$H_0$ plane using the ET data. Lower panel: Two-dimensional marginalized contours (68.3\% and 95.4\% confidence level) in the $\Omega_{\rm m}$--$H_0$ plane using the ET2CE data.}
\end{figure}

\begin{table*}[!htb]
	\caption{The absolute (1$\sigma$) and relative errors of the cosmological parameters in the $\Lambda$CDM model using four typical mass distributions of NSs of ET and ET2CE data in the optimistic scenario for the FOV.}
	\label{tab8}
	\setlength{\tabcolsep}{2.15mm}
	\renewcommand{\arraystretch}{1.5}
	\begin{center}{\centerline{
				\begin{tabular}{|c|c|m{2.3cm}<{\centering}|m{2.3cm}<{\centering}|m{2.3cm}<{\centering}|m{2.3cm}<{\centering}|}
					\hline   
					\multicolumn{2}{|c|}{NS mass distribution}   &Galactic BNSs   &Galactic NSs &POWER model &PEAK model \\ \hline
					\multirow{4}{*}{ET}
					&$\sigma(\Omega_{\rm m})$       &$0.013$   &$0.012$    &$0.011$     &$0.011$       \\
					&$\sigma(H_0)$                            &$0.14$     &$0.13$      &$0.13$       &$0.11$         \\
					&$\ve(\Omega_{\rm m})$             &$4.18\%$  &$3.86\%$&$3.54\%$  &$3.54\%$    \\
					&$\ve(H_0)$                                  &$0.207\%$ &$0.192\%$ &$0.192\%$ &$0.163\%$      \\ \hline
					\multirow{4}{*}{ET2CE}
					&$\sigma(\Omega_{\rm m})$          &$0.0052$   &$0.0049$ &$0.0048$&$0.0047$       \\
					&$\sigma(H_0)$                               &$0.044$  &$0.041$  &$0.040$    &$0.039$        \\
					&$\ve(\Omega_{\rm m})$                 &$1.67\%$   &$1.58\%$  &$1.55\%$   &$1.51\%$    \\
					&$\ve(H_0)$                                     &$0.065\%$ &$0.061\%$&$0.059\%$&$0.058\%$    \\    \hline				
		\end{tabular}}}
	\end{center}
\end{table*}

In Fig.~\ref{fig15}, we show the distributions of luminosity distance uncertainty $\Delta d_{\rm L}/d_{\rm L}$ with respective to the redshifts for GW standard sirens using four mass distributions of NSs. We find that $\Delta d_{\rm L}/d_{\rm L}$ and redshift distributions exhibit only slight differences across the four NS mass distributions. In Fig.~\ref{fig16}, we show the constraint results in the $\Omega_{\rm m}$--$H_0$  plane for the $\Lambda$CDM model using four mass distributions of NSs of ET and ET2CE. The detailed results are given in Table~\ref{tab8}. We can see that the four NS mass distributions give similar constraint results, although the constraint results of the Galactic BNS mass distribution are slightly worse than those of the other NS mass distributions (the errors given by the Galactic BNS mass distribution are slightly higher than those of the other NS mass distributions). This means that the mass distributions of NSs have less impact on the cosmological analysis.

\subsection{Comparison of the number of standard sirens and GW detection strategies}

The impact of GWs on cosmological parameter constraints primarily arises from the number of standard sirens and the error of luminosity distance. For the same source, the errors measured by different GW detection strategies are different. In this subsection, we briefly discuss the impact of these two factors on cosmological analysis. Note that the following discussions are based on the $\Lambda$CDM model.

\begin{figure}[htbp]
	\includegraphics[width=0.9\linewidth,angle=0]{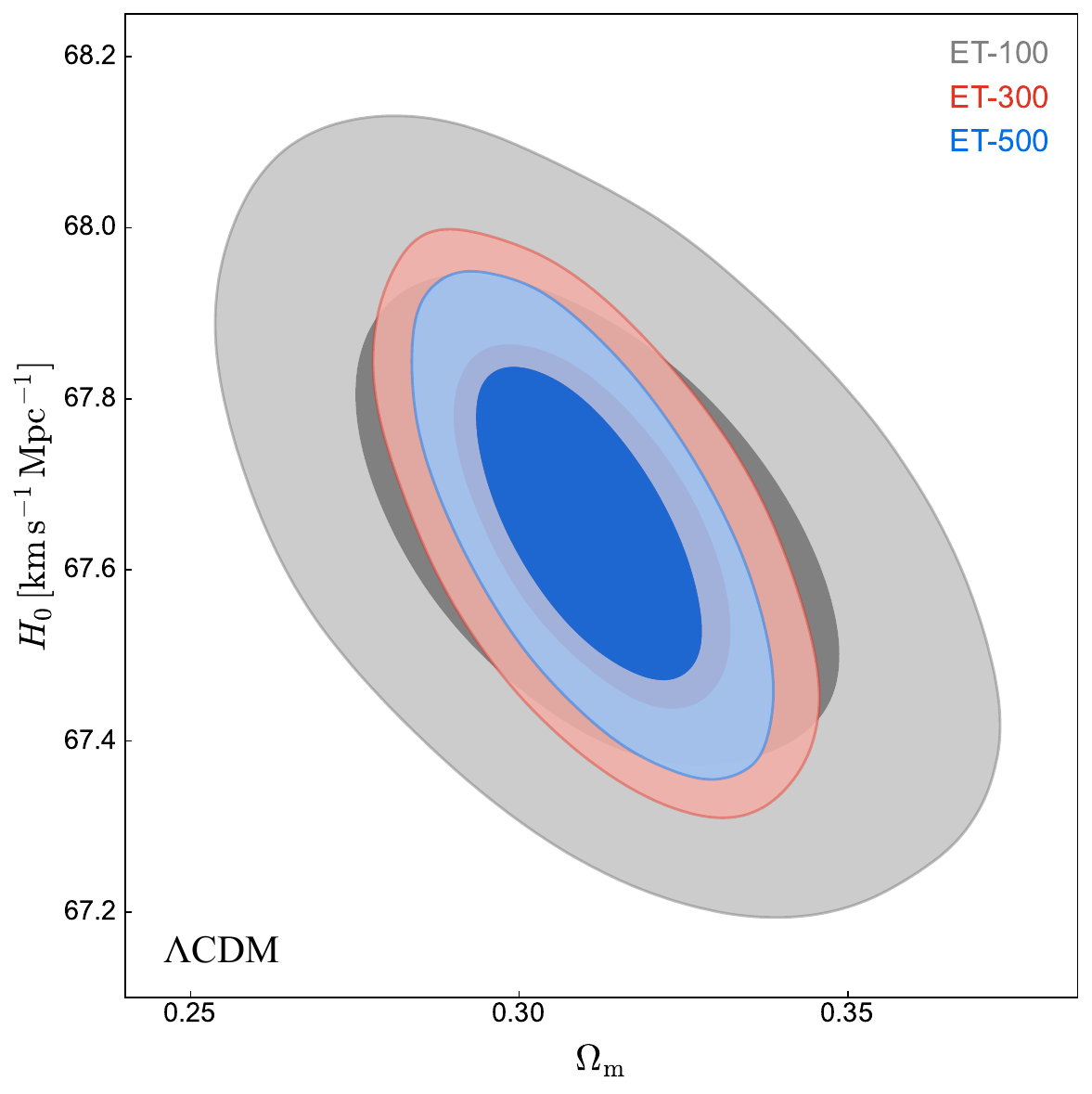}
	\caption{\label{fig17} Two-dimensional marginalized contours (68.3\% and 95.4\% confidence level) in the $\Omega_{\rm m}$--$H_0$ plane for the $\Lambda$CDM model using 100, 300 and 500 standard sirens of ET, respectively.}
\end{figure}

\begin{table}[!h]
	\caption{The absolute (1$\sigma$) and relative errors of the cosmological parameters in the $\Lambda$CDM model using 100, 300 and 500 standard sirens of ET, respectively.}
	\label{tab9}
	\setlength{\tabcolsep}{2mm}
	\renewcommand{\arraystretch}{1.5}
	\begin{center}
		\scalebox{0.85}{\centerline{\begin{tabular}{|p{1.5cm}<{\centering}|p{2cm}<{\centering}|p{2cm}<{\centering}|p{2cm}<{\centering}|}
					\hline
					Number & 100 & 300 & 500\\
					\hline
					$\sigma(\Omega_{\rm m})$& $0.024$ &  $0.014$ & $0.011$\\
					\hline
					$\sigma(H_0)$& $0.19$ & $0.14$ & $0.12$\\
					\hline
					$\varepsilon(\Omega_{\rm m})$& $7.69\%$ & $4.50\%$ & $3.54\%$\\
					\hline
					$\varepsilon(H_0)$& $0.281\%$ &  $0.207\%$ & $0.177\%$\\
					\hline
		\end{tabular}}}
	\end{center}
\end{table}

\begin{figure}[htbp]
	\includegraphics[width=0.9\linewidth,angle=0]{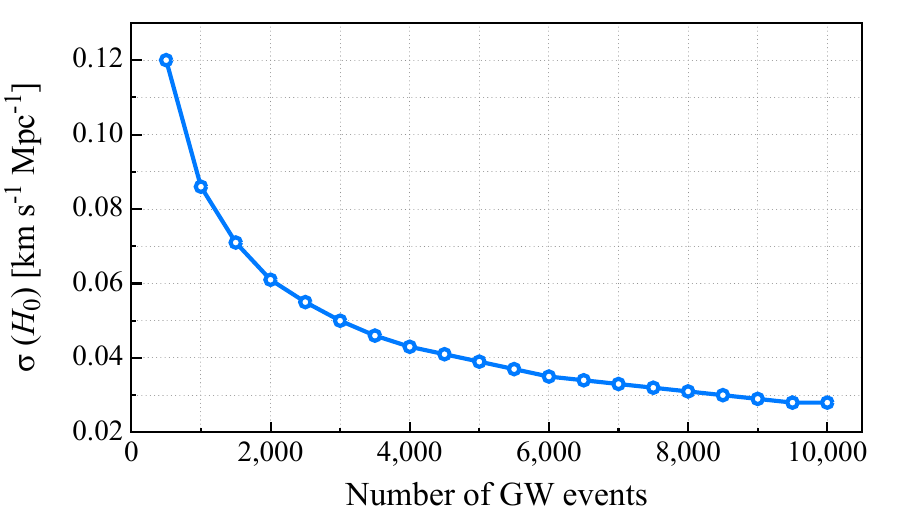}
	\includegraphics[width=0.9\linewidth,angle=0]{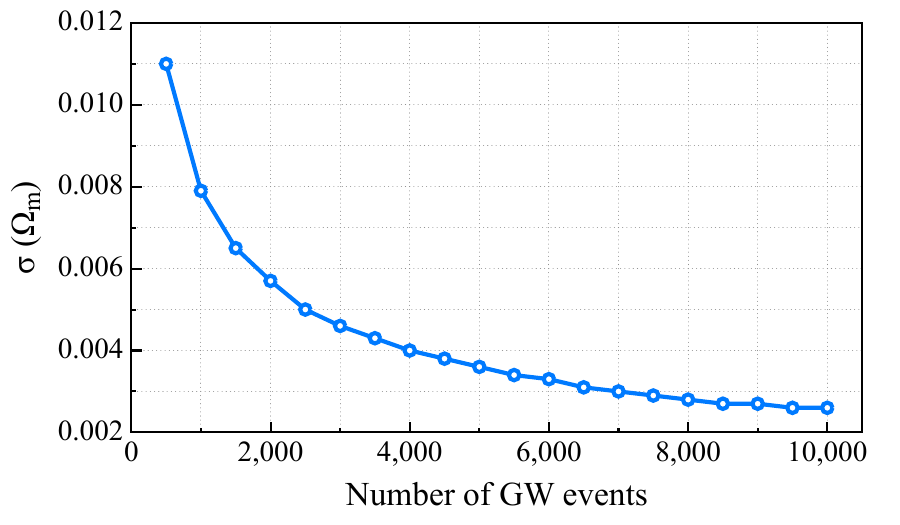}
	\caption{\label{fig18} The absolute (1$\sigma$) errors of the cosmological parameters in the $\Lambda$CDM model using different numbers of standard sirens of ET. The upper panel shows the results of $H_0$ and the lower panel shows those of $\Omega_{\rm m}$.}
\end{figure}

We consider a specific case of ET as a concrete example to analyze the impact of the number of standard sirens. In Fig.~\ref{fig17}, we show the  constraint results in the $\Omega_{\rm m}$--$H_0$ plane for the $\Lambda$CDM model using 100, 300 and 500 standard sirens of ET, whose redshift distributions are proportional to those in Fig.~\ref{fig5}. The detailed results are given in Table~\ref{tab9}. As can be seen, ET with 100 standard sirens gives the worst constraint results. Concretely, the constraint precisions of the parameters $\Omega_{\rm m}$ and $H_0$ from 300 standard sirens could be improved by 41.7\% and 26.3\% than those of 100. Conversely, the constraint results from 500 standard sirens are slightly better than those of 300. To further analyze the impact of the number of standard sirens on cosmological analysis, we increased the number of standard sirens used to constrain cosmological parameters, although this prediction is overly optimistic. The results are shown in Fig.~\ref{fig18}. This results show that the number of standard sirens has an important impact on cosmological estimations, but once it reaches a certain level, further improvement in cosmological parameters becomes insignificant.

\begin{figure}[htbp]
	\includegraphics[width=0.9\linewidth,angle=0]{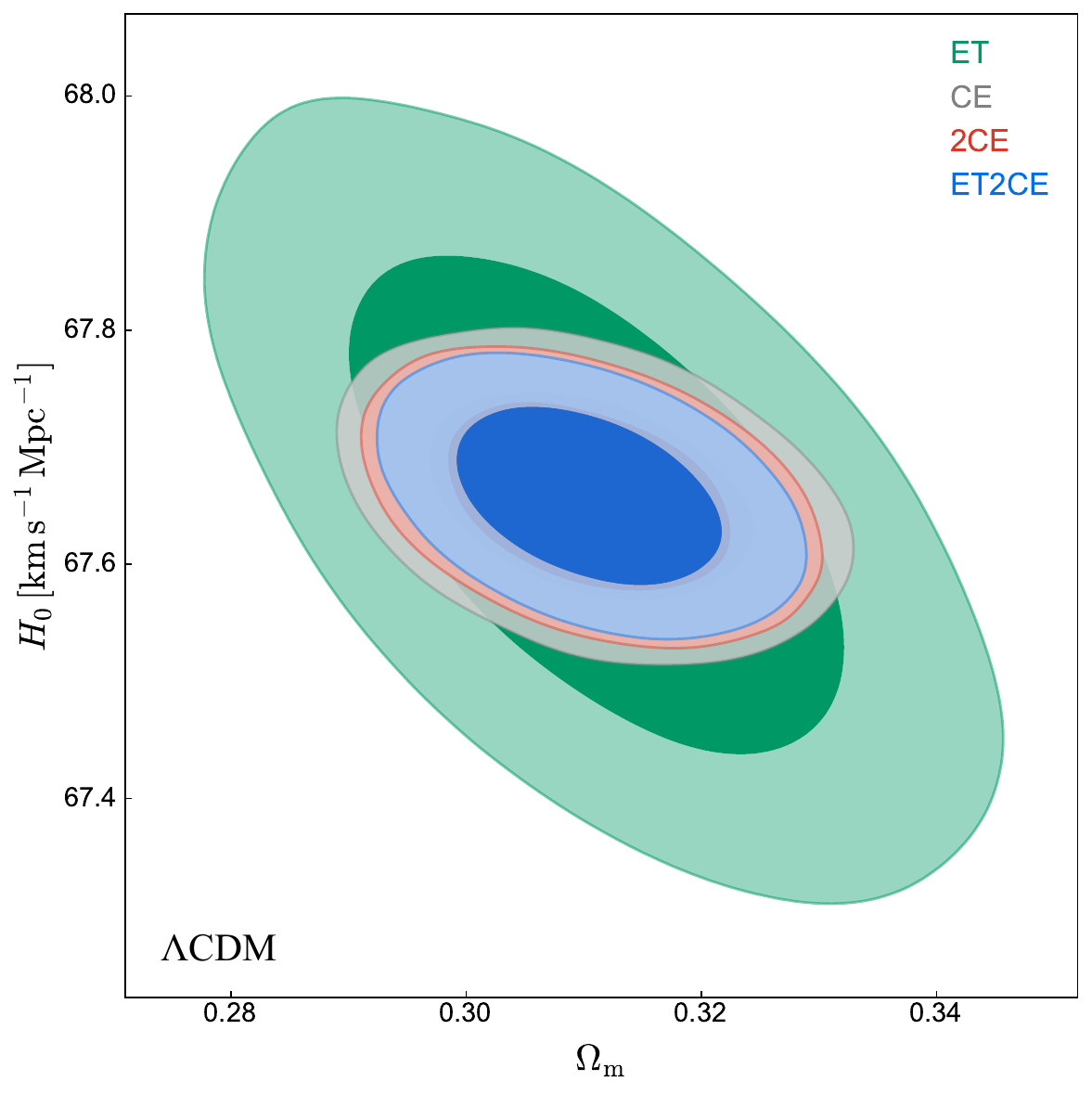}
	\caption{\label{fig19} Two-dimensional marginalized contours (68.3\% and 95.4\% confidence level) in the $\Omega_{\rm m}$--$H_0$ plane for the $\Lambda$CDM model using 300 standard sirens of ET, CE, 2CE, and ET2CE, respectively.}
\end{figure}

\begin{table}[!h]
	\caption{The absolute (1$\sigma$) and relative errors of the cosmological parameters in the $\Lambda$CDM model using 300 standard sirens of ET, CE, 2CE, and ET2CE, respectively.}
	\label{tab10}
	\setlength{\tabcolsep}{1.5mm}
	\renewcommand{\arraystretch}{1.5}
	\begin{center}
		\scalebox{0.85}{\centerline{\begin{tabular}{|p{1.5cm}<{\centering}|p{1.5cm}<{\centering}|p{1.5cm}<{\centering}|p{1.5cm}<{\centering}|p{1.5cm}<{\centering}|}
					\hline
					Detector & ET & CE & 2CE &ET2CE\\
					\hline
					$\sigma(\Omega_{\rm m})$& $0.0140$ &  $0.0089$ & $0.0080$ & $0.0074$\\
					\hline
					$\sigma(H_0)$& $0.140$ & $0.059$ & $0.053$& $0.050$\\
					\hline
					$\varepsilon(\Omega_{\rm m})$& $4.50\%$ &$2.87\%$ & $2.58\%$ & $2.38\%$\\
					\hline
					$\varepsilon(H_0)$& $0.207\%$ &  $0.087\%$ & $0.078\%$& $0.074\%$\\
					\hline
		\end{tabular}}}
	\end{center}
\end{table}

In Fig.~\ref{fig19}, we show the constraint results in the $\Omega_{\rm m}$--$H_0$ plane for the $\Lambda$CDM model using 300 standard sirens of ET, CE, 2CE, and ET2CE, whose redshift distributions are proportional to those in Fig.~\ref{fig5}. The detailed results are shown in Table~\ref{tab10}. We can see that ET gives the worst constraint results. Compared with ET, the constraint precisions of the parameters $\Omega_{\rm m}$ and $H_0$ from CE could be improved by 36.4\% and 57.9\%.  Conversely, the 2CE and ET2CE’s constraining capabilities on the cosmological parameters are slightly better than those of CE. Therefore, with the same number of standard sirens, different GW detection strategies have significant impacts on cosmological analysis. CE, 2CE, and ET2CE give similar constraint results, which are much better than those of ET.

\section{Conclusion}\label{sec6}

In this work, we show the potential of the GW standard sirens from the 3G GW detectors in constraining cosmological parameters. We explore the synergy between 3G GW detectors and GRB detector THESEUS-like telescope for the multi-messenger observations. We consider four GW observation strategies, i.e., ET, CE, the 2CE network, and the ET2CE network. Five cosmological models, $\Lambda$CDM, $w$CDM, $w_0w_a$CDM, I$\Lambda$CDM, and I$w$CDM, are considered to perform cosmological analysis. Moreover, we consider the optimistic (assuming all the detected short GRBs could determine redshifts) and realistic (assuming 1/3 of the detected short GRBs could determine redshifts) cases for FOV to make the multi-messenger analysis.

We first predict the expected detection rates of the GW-GRB events based on the optimistic and realistic cases. For the detector network, the expected number of GW-GRB is almost double compared to the single ET observatory. About $\mathcal{O}(100)$ GW-GRB events could be detected based on the 10-year observation. Moreover, the detected GW-GRB events have $z<3.5$ and $\iota<15^\circ$.

We find that GW gives quite tight constraints on the Hubble constant, with precisions from 0.345\% to 0.065\%. However, GW gives loose constraints on the other cosmological parameters in both optimistic and realistic scenarios. Fortunately, since GW has an advantage in measuring the absolute luminosity distance, GW can break the cosmological parameter degeneracies generated by other EM observations, thus improving the measurement precisions of cosmological parameters. When combining CBS with ET2CE, the constraint precision of the EoS parameter of dark energy $w$ can reach 1.26\%, which is close to the standard of precision cosmology. With the addition of ET2CE to CBS, the constraints on cosmological parameters can be improved by 34.2\%--94.9\%. We can conclude that (i) the synergy between 3G GW detectors and THESEUS could detect $\mathcal{O}(100)$ multi-messenger events based on the 10-year observation; (ii) GW can provide rather precise measurement on the Hubble constant with a precision of 0.065\%, but poor at measuring the other cosmological parameters; (iii) GW can significantly break the cosmological parameter degeneracies generated by the other EM observations and the combination of them is expected to precisely measure dark energy. It is worth expecting that GW standard sirens from the 3G GW detectors can help make arbitration for the Hubble tension and explore the fundamental nature of dark energy.

\begin{acknowledgments}
We thank Ling-Feng Wang, Ming Zhang, Peng-Ju Wu, and Shuang-Shuang Xing for helpful discussions. This work was supported by the National SKA Program of China (Grants Nos. 2022SKA0110200 and 2022SKA0110203), the National Natural Science Foundation of China (Grants Nos. 11975072, 11875102, and 11835009), and the 111 Project (Grant No. B16009).
\end{acknowledgments}
\bibliography{gw_grb}
\end{document}